\documentclass[usegraphix, useAMS, fleqn, usenatbib]{mnras}

\usepackage{amsmath}
\usepackage{amssymb}

\usepackage{xcolor}
\usepackage{graphicx}
\usepackage{dcolumn}
\usepackage{bm}
\usepackage{hyperref}
\hypersetup{colorlinks, allcolors={blue}}
\usepackage{soul}
\usepackage{siunitx}
\usepackage{longtable}

\DeclareRobustCommand*{\unit}[2][]{%
  \def\@unit@value{#1}%
  \ifx\@unit@value\@empty\else#1\,\fi
  \ifmmode\mathrm{#2}\else#2\fi
}

\makeatletter
\newwrite\figdimsfile
\AtEndDocument{%
  \immediate\openout\figdimsfile=figure_dims.cfg
  \immediate\write\figdimsfile{textwidth_pt=\the\textwidth}
  \immediate\write\figdimsfile{columnwidth_pt=\the\columnwidth}
  \immediate\write\figdimsfile{fontsize_pt=\f@size}
  \immediate\closeout\figdimsfile
}
\makeatother

\title[]{The presence and order of QCD phase transitions from gravitational-wave observations of neutron star mergers}

\author[Lasky et al. ]{
Paul D. Lasky,$^{1,2,3}$
Alexander Clevinger,$^{4}$
Andr\'e G. da Silva,$^{5}$ 
\newauthor
\,\,Ricardo L. S. Farias,$^{5,4}$
and Veronica Dexheimer,$^{4,6}$
\\
$^{1}$School of Physics and Astronomy, Monash University, VIC 3800, Australia\\
$^{2}$OzGrav: The ARC Centre of Excellence for Gravitational-wave Discovery, Australia\\
$^{3}$Department of Physics and Astronomy, University of British Columbia, Vancouver, British Columbia, V6T1Z4, Canada\\
$^{4}$Center for Nuclear Research, Department of Physics, Kent State University, Kent, 44242, OH, USA\\
$^{5}$Departamento de F\'{i}sica, Universidade Federal de Santa Maria, 97105-900 Santa Maria, RS, Brazil.\\
$^{6}$CERCA/ISO, Department of Physics, Case Western Reserve University, Cleveland, Ohio 44106, USA.
}

\begin{document}
\maketitle

\begin{abstract}
Gravitational-wave observations of neutron star mergers allows us to probe the dense portion of the nuclear equation of state. We develop a formalism for testing the existence and order of phase transitions in the cores of neutron stars using only gravitational-wave observations of their inspiral. We test our formalism on data from the first binary neutron star mergers GW170817 and GW190425, and use synthetic injections assuming the next and uber-next upgrades of the LIGO observatories, A+ and A$^\sharp$, as well as with Cosmic Explorer. We show that populations of mergers in either A+ or A$^\sharp$ will be able to measure the existence of a first-order phase transition in observations lasting less than one PhD (less than half a US PhD!). On the other hand, with only the inspiral, we cannot distinguish populations of neutron stars with no phase transition from those with second- or third-order phase transitions or with first-order phase transitions with relatively large transition densities. 
\end{abstract}

\section{Introduction}
Gravitational-wave astronomy offers a unique way to probe nuclear physics in neutron star cores---extreme environments that remain beyond the reach of terrestrial experiment, electromagnetic observation, and first-principles theory. Crucially, this is precisely where novel physics may emerge, including states in which quarks become deconfined. Here, we evaluate whether gravitational-wave signals from inspiralling binaries can identify such phase transitions in cold neutron stars, and whether we can constrain their order.

The outer core of neutron stars is almost certainly made of hadrons: subatomic particles made of two or more quarks bound by the strong force. At extreme densities, those hadrons become so tightly squeezed that their Fermi energy rises dramatically. Deconfining the bound quarks allows them to lower their Fermi energy by distributing momentum across colour and flavour degrees of freedom. This represents a phase transition from a purely hadronic state to pure quark matter. Perturbative quantum chromodynamics (QCD) calculations at asymptotically high energies, far exceeding conditions found in neutron stars, show that these deconfined quark states must exist~\citep[e.g.,][]{collins75,freedman77,fraga05,ghiglieri20}, although it is unclear where the transition happens inside neutron stars~\citep[e.g.,][]{shirke23}; for a review of constraints, see~\citet{kumar24}.

Phase diagrams often depict a first-order phase transition between hadronic and quark matter, although this has not been experimentally verified. But the so-called \textit{hyperon puzzle}~\citep[e.g., see][and references therein]{bednarek12,chatterjee16,burgio21} motivates broader thinking about the order of the phase transition. Hyperons are expected to appear in neutron stars due to high chemical potentials, however their existence adds further degrees of freedom to the system, which softens the equation of state, and in turn makes it difficult to generate neutron stars with masses consistent with the most massive observed in nature~\citep{cromartie20,fonseca21}. One possible resolution to the hyperon puzzle is that the deconfined quark state appears through phase transition more similar to a smooth crossover, or alternatively a higher-order phase transition, between the hadronic and quark phases~\citep{masuda16,baym18}. The smoother transition acts to stiffen the equation of state around the density at which hyperons become necessary, effectively allowing higher masses consistent with observations. 

Gravitational waves from merging neutron stars offer a method to probe the equation of state, and hence potentially the existence or not of a phase transition~\citep[e.g.,][]{chatziioannou20,pang20,essick23}. This is true both in the cold state during the inspiral, and potentially the hot state with post-merger remnants~\citep[e.g.,][]{most19,bauswein19,sarin21}. The observation of binary neutron star mergers started with a bang in 2017~\citep{abbott17_GW170817detection,abbott17_GW170817MMA}, but has reduced to a whimper in the subsequent decade. This should be about to change. The LIGO observatories~\citep{aasi15} are hoping to upgrade to so-called A$+$ sensitivity~\citep{abbott18_Aplus} in the coming few years, with binary neutron star mergers predicted to be observed at a rate of $\unit[17^{+44}_{-14}]{yr^{-1}}$. The next upgrade beyond that, so-called $A^{\sharp}$, has a baseline prediction of $\unit[69^{+160}_{-51}]{yr^{-1}}$~\citep{sun26}. These numbers only include LIGO Hanford and Livingston observatories, and says nothing of the additional sensitivity  Virgo~\citep{acernese15}, KAGRA~\citep{akutsu21}, and/or LIGO India~\citep{iyer11} will add. 

The increased event rate of neutron star mergers may not, in and of itself, be the key to getting precision details about the equation of state. Instead, it is the loudest of events that provide the most information~\cite[e.g., see][]{lackey15,hernandezvivanco19}. A GW170817-like event in the A+ era will have a signal-to-noise ratio (SNR) $\approx162$, and $\approx206$ for A$^{\sharp}$. With that said, if we have learnt anything over the last ten years, it is that GW170817-like events are relatively rare. Back of the envelope, if we assume it was a one-in-ten-year event, then we would still expect a couple of neutron star mergers per year with SNR greater than that of GW170817 in A+, and more than five per year in $A^\sharp$. 

The above says nothing about third-generation observatories Einstein Telescope~\citep{punturo10} and Cosmic Explorer~\citep{abbott17_CE}. Those instruments anticipate upward of 100,000 neutron star mergers per year~\citep[e.g.,][]{maggiore20,evans21,branchesi23}, with dozens per year well above an SNR of 100.

And so it is with moderate confidence that we embark on a journey of discovery to understand what can be learned about the possible existence and nature of phase transitions from future gravitational-wave measurements. In this Paper, we develop a framework for measuring the existence and order of a potential phase transition both in single  neutron star merger measurements and in a full population of mergers.

\section{Phase transitions in neutron stars}\label{sec:phase_transitions}

The order of a phase transition is defined by the lowest-order derivative of the free energy that is discontinuous as the system crosses the transition. 
In a first-order phase transition, a physical quantity that is a first derivative of the free energy---such as density---changes discontinuously. In a second-order phase transition, this first derivative varies continuously across the transition, and the second derivative of the free energy---such as a compressibility---is discontinuous. More generally, an $m$th-order phase transition is one for which every derivative of the free energy up to order $m-1$ is continuous, with the first discontinuity appearing in the $m$th derivative. For the dense matter making up neutron stars, the relevant free energy is the pressure $P$ as a function of the baryon chemical potential $\mu_B$: its first derivative is the baryon number density, $n_B=\partial P/\partial\mu_B$, and its higher derivatives are the so-called baryon susceptibilities $\chi_n^B=\partial^nP/\partial\mu_B^n$. 

\begin{figure}
    \centering
    \includegraphics[width=\columnwidth]{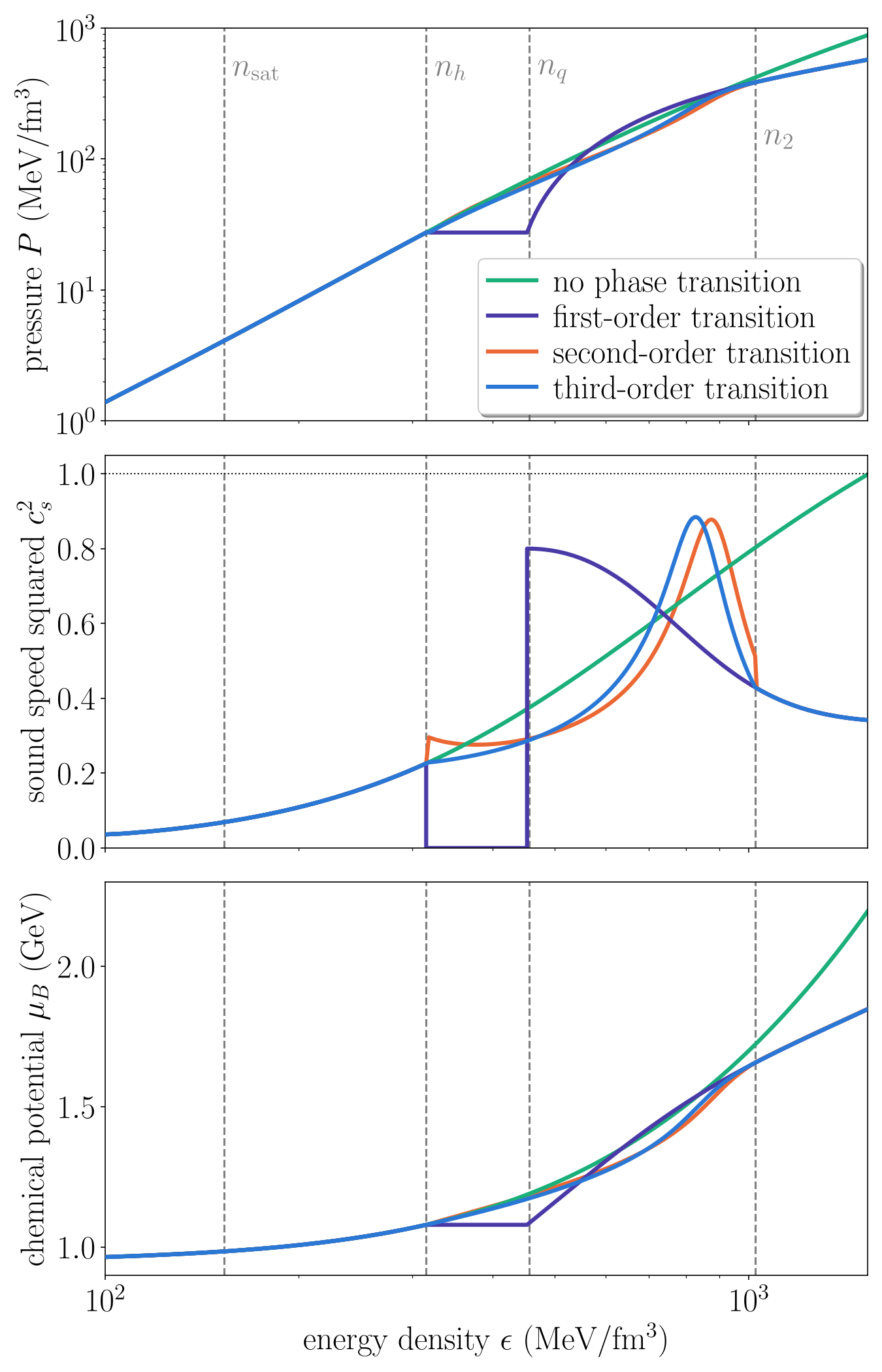}
    \caption{Illustrative example of four equations of state built from the same underlying baseline equation of state SK272 (green curve). That equation of state has no phase transition, but we manually add a first- (purple), second- (orange), and third-order (blue) phase transition as described in the text. The panels show pressure (top), sound speed squared $c_s^2$ (middle), and baryon chemical potential $\mu_B$ (bottom) as functions of energy density. The dashed grey lines mark nuclear saturation density $n_{\rm sat}$ and the various transition densities described in the text.}
    \label{fig:soundspeed_pressure}
\end{figure}

\begin{figure*}
    \centering
    \includegraphics[width=0.95\textwidth]{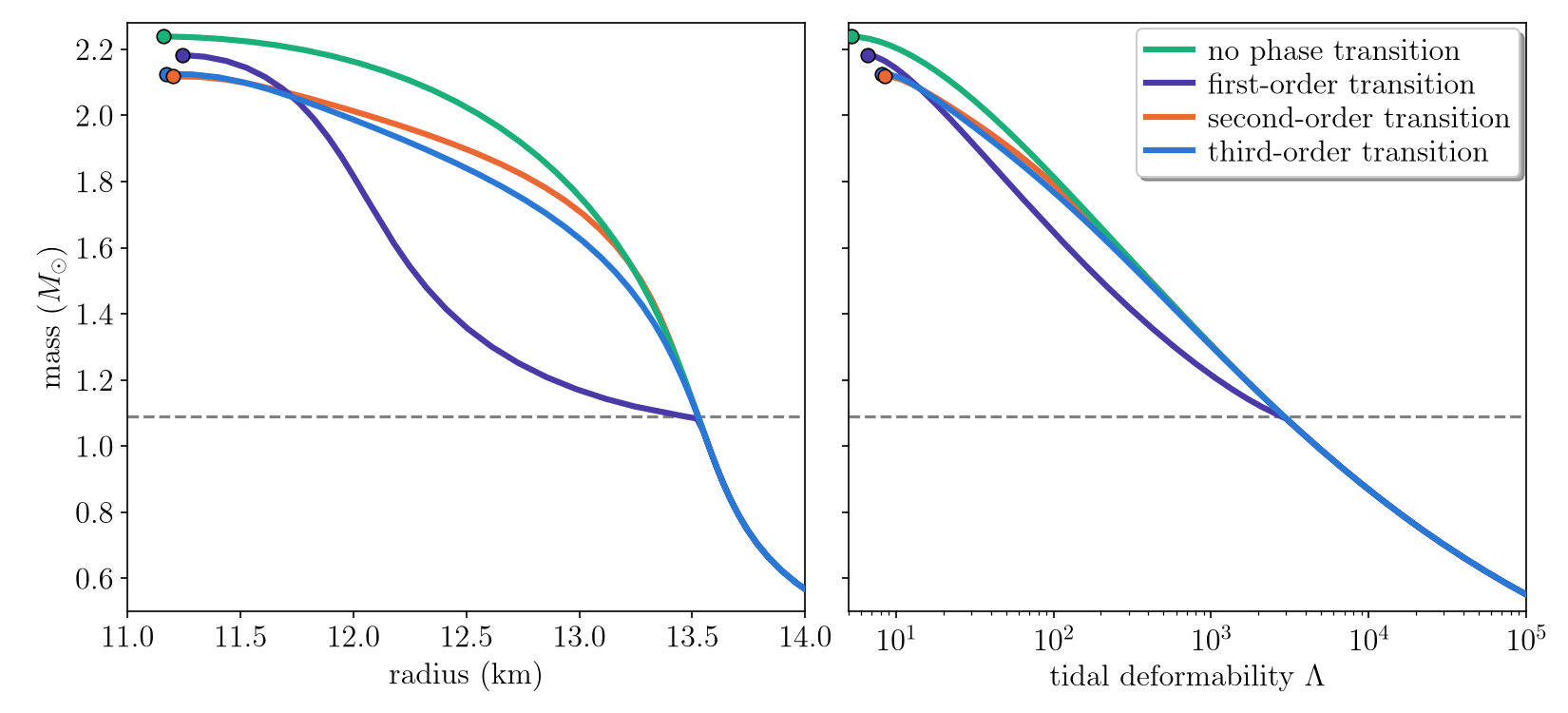}
    \caption{Mass--radius (left) and mass--tidal-deformability (right) relations for the same four equations of state as Fig.~\ref{fig:soundspeed_pressure}: no phase transition, and first-, second-, and third-order phase transitions. The dashed grey line marks the stellar mass at which a star's central density first reaches the transition density. Filled circles mark each curve's maximum mass.}
    \label{fig:massradiuslambda}
\end{figure*}

Figures~\ref{fig:soundspeed_pressure} and~\ref{fig:massradiuslambda} show illustrative examples of the phase transitions we are using in this work. In Figure~\ref{fig:soundspeed_pressure}, we show pressure $P$ (top panel), sound speed squared $\partial P/\partial\epsilon=c_s^2$ (middle panel; in units where the speed of light $c=1$), and baryon chemical potential $\mu_B$ (bottom panel) as functions of energy density $\epsilon$. The base nuclear equation of state\footnote{All base nuclear equations of state used are accessed through the CompOSE database~\citep{typel15,typel22}.}, shown in green, is the SK272 Skyrme energy-density functional~\citep{agrawal03} that has nuclear matter incompressibility of $K_{\rm nm}\approx\unit[272]{MeV}$.

Shown in purple, orange, and blue, are respectively the equations of state with added first-, second-, and third-order phase transitions. We give these three equations of state identical descriptions to the base nuclear equation of state below the first transition point at baryon number density $n_h$. That is, at baryon number density $n<n_h$, all four equations of state are identical and described by purely hadronic matter.  We also give our three equations of state with phase transitions identical descriptions above some critical density such that they follow a shared quark-matter branch (for first-order phase transitions this is $n_q$, and for second- and third-order this is $n_2\geq n_q$, which corresponds to another phase transition of the same order as the previous one). For each base nuclear equation of state, there are a number of parameters we use to describe the various transition regions, resulting in thousands of candidate equations of state at each phase-transition order.

We ensure that all equations of state containing quark matter (i.e., all those with any order of phase transition) satisfy the asymptotic conformal limit of $c_s^2=1/3$ expected for deconfined quark matter at asymptotically high density~\citep{annala20}. This is a different prescription from \citet{clevinger26}, who instead fixes the high-density boundary condition such that it smoothly matches onto the chiral mean field CMF-7 hybrid equation of state~\citep{clevinger22}. It is not possible to use this upper boundary for a first-order phase transition which requires the hadronic and quark branches to have the exact same pressure and chemical potential at the transition. That is, we cannot generically find a crossing point between an arbitrary hadronic equation of state and the quark table equation of state used in~\citet{clevinger26}, and we would therefore not be able to do a direct comparison between first-order transitions and second- and third-order transitions. 

We include four free parameters for our first-order transition curves, following a modified version of the constant-sound-speed construction~\citep{alford13, chatziioannou20}, which are $\left\{n_h,\,n_q,\,c_{s,{\rm peak}},\,\Delta\right\}$. Here, $n_h$ is the first transition baryon number density which, for Figure~\ref{fig:soundspeed_pressure}, we set at twice the nuclear saturation density $n_h=2n_{\rm sat}=\unit[0.32]{fm^{-3}}$. We define the number density at which the transition ends and the equation of state is therefore quark matter as $n_q$, which for Figure~\ref{fig:soundspeed_pressure} and the first-order phase transition is $n_q=1.4n_h$. The peak sound speed (which is the value of the sound speed at $n_q$) the square of which is $c_{s,\rm peak}^2=0.8$ in Figure~\ref{fig:soundspeed_pressure}. The final parameter is a relaxation parameter $\Delta$ that dictates how quickly the sound-speed overshoot relaxes back to $c_s^2=1/3$. We implement $\Delta$ according to
\begin{align}
    c_s^2(\epsilon) = \frac{1}{3} + \left(c_{s,\rm peak}^2 - \frac{1}{3}\right)\mathrm{sech}^2\left(\frac{\epsilon-\epsilon_q}{\Delta}\right),
\end{align}
where $\epsilon_q$ is the energy density corresponding to $n_q$, the point at which $c_s^2$ jumps discontinuously from zero up to its peak value $c_{s,\rm peak}^2$. In Figure~\ref{fig:soundspeed_pressure}, this is chosen as $\Delta=\unit[400]{MeV fm^{-3}}$.

The second- and third-order phase transition constructions largely follow~\citet{clevinger26}, inspired by~\citet{kojo15}, with a change to the large-$\epsilon$ behaviour. In particular, we enforce the speed of sound asymptotes to the high-density quark limit of $c_s^2=1/3$ using the $\Delta$ construction outlined above, instead of matching with the original equation of state. The second-order phase transition curve inherits the four parameters from the first-order transition (which, for Figure~\ref{fig:soundspeed_pressure}, retain the same values as the first-order curve), and includes three new parameters $\left\{n_2,\,\chi_2^{B,q},\,\chi_2^{B,h}\right\}$. Here, $n_2\geq n_q$ is the density at which the fitted curve hands off to the quark matter branch, which for Figure~\ref{fig:soundspeed_pressure} is $n_2=\unit[0.85]{fm^{-3}}$. The two baryon susceptibilities $\chi_2^B$ allow for a change in the size of the discontinuity at the transition region. There are two of these parameters evaluated at the hadronic and quark boundaries, which for Figure~\ref{fig:soundspeed_pressure} respectively take on values $\chi_2^{B,h}=\chi_2^{B,q}=\unit[0.001]{fm^{-3}\,MeV^{-1}}$.

Finally, our third-order transition equations of state have two \textit{fewer} free parameters than second-order curves because they enforce a smooth transition at the $n_h$ and $n_2$ boundaries, which forces values of $\chi_2^{B,h}$ and $\chi_2^{B,q}$ to be equal on either side of the corresponding boundary.  

For our full set of candidate equations of state, we establish a grid of parameters resulting in a large number of candidate equations of state for each base nuclear equation of state. Table~\ref{tab:transition_grids} provides details of the parameter ranges we use for our grid of candidate equations of state.

The lower bound of our onset-density range is physically motivated. For our baseline equations of state, a star whose central density first reaches $n_h=2\,n_{\rm sat}=\unit[0.32]{fm^{-3}}$ has a mass of only $\approx1.0\,M_\odot$ (median over baselines), falling to $\approx0.6\,M_\odot$ at $n_h=1.5\,n_{\rm sat}$. Onset densities $n_h\lesssim2\,n_{\rm sat}$ therefore have essentially every observable neutron star be a hybrid star. We therefore do not extend our grid below $n_h=1.5\,n_{\rm sat}$. The upper bound of $n_h$ is chosen such that the transition is reached only in stars more massive than $\approx1.7\,M_\odot$ (median over baselines), so typical neutron stars of $\approx1.4\,M_\odot$ remain purely hadronic.

Our range of $n_q/n_h$ similarly spans the physically interesting regime. Because the first-order segment is at constant pressure and chemical potential, the density jump corresponds to a transition strength $\Delta\varepsilon/\varepsilon_{\rm t}=(n_q/n_h-1)(1+P_{\rm t}/\varepsilon_{\rm t})$, which for our baselines ranges from $\approx0.2$ at $n_q/n_h=1.2$, a weak transition of the kind found to be difficult to detect~\citep{chatziioannou20}, to $\approx1.1$ at $n_q/n_h=2$. The upper part of this range exceeds the Seidov criterion, $\Delta\varepsilon/\varepsilon_{\rm t}>1/2+3P_{\rm t}/(2\varepsilon_{\rm t})$, above which the hybrid branch becomes unstable at the onset of the transition~\citep[e.g.,][]{alford13}; for our baselines this threshold lies at $n_q/n_h\approx1.55$--$1.65$.

\begin{table}
    \centering
    \caption{Parameter grids for all equations of state with phase transitions used in this work. Each parameter is sampled at $N$ points with linear (endpoint-inclusive) spacing. The final column shows the transition orders for which the respective parameters are relevant. We add the constraint that $n_2 \geq n_q$, and use 111 base nuclear equations of state, which together results in 1,110,000 first-order, 98,901,000 second-order, and 10,989,000 third-order candidate equations of state.}
    \label{tab:transition_grids}
    \begin{tabular}{lcccc}
        \hline
        Parameter & Min & Max & $N$ & Order\\
        \hline
        $n_h$ [$\mathrm{fm}^{-3}$] & $0.24$ & $0.48$ & $10$ & \{1,2,3\} \\
        $n_q/n_h$ & $1.2$  & $2.0$  & $10$ & \{1,2,3\}\\
        $c_{s,\mathrm{peak}}^{2}$ & $0.5$ & $0.95$ & $10$ & \{1,2,3\}\\
        $\Delta$ [$\mathrm{MeV/fm^{3}}$] & $100$ & $600$ & $10$ & \{1,2,3\}\\
        $n_2$ [$\mathrm{fm}^{-3}$] & $0.85$ & $1.3$ & $10$ & \{2,3\}\\
        $\chi_2^{B,h}$ [$\mathrm{fm^{-3}\,MeV^{-1}}$] & $4\times10^{-4}$ & $3\times10^{-3}$ & $3$ & \{2\}\\
        $\chi_2^{B,q}$ [$\mathrm{fm^{-3}\,MeV^{-1}}$] & $4\times10^{-4}$ & $3\times10^{-3}$ & $3$ & \{2\}\\
        \hline
    \end{tabular}
\end{table}

We construct each equation of state using the open-source~\textsc{Percolation-Project}~\citep{clevinger26}, which we extend with the first-order contribution described above. This allows us to calculate macroscopic neutron star quantities for each of our representative equations of state. Figure~\ref{fig:massradiuslambda} shows the macroscopic quantities for our illustrative four equations of state shown in Figure~\ref{fig:soundspeed_pressure}. The left panel shows the neutron star mass as a function of the radius for each equation of state, as calculated by solving the Tolman-Oppenheimer-Volkoff equations of hydrostatic equilibrium. The horizontal line in that plot represents the neutron star mass at which the central density of the star reaches $n_h$. Below that point, all equations of state yield the same macroscopic quantities because the first transition density is not realised in the star. Above that point, we see a non-smooth change to the mass-radius curve for the first-order transition equation of state due to the discontinuous energy density. On the other hand, the second- and third-order transition equations of state deviate from the base nuclear equation of state smoothly. 

The base nuclear equation of state used here has a maximum neutron star mass, shown as the filled circle for each equation of state, of $M_{\rm TOV}=2.24\,M_\odot$. Adding the phase transitions tends to decrease the maximum mass, with the first-, second-, and third-order equations of state having $M_{\rm TOV}=2.18\,M_\odot$, $2.12\,M_\odot$ and $2.13\,M_\odot$ respectively.\footnote{This decrease is not generic: a transition followed by a sufficiently stiff quark branch can instead increase the maximum mass, particularly for baselines with a low $M_{\rm TOV}$, such as the baseline used by \citet{clevinger26}. We retain such equations of state; approximately 30 per cent of the candidates in our grid have a larger $M_{\rm TOV}$ than their baseline.}

The right-hand panel of Figure~\ref{fig:massradiuslambda} shows the dimensionless tidal deformability 
\begin{align}
    \Lambda=\frac{2}{3}k_2\left(\frac{c^2 R}{GM}\right)^5,
\end{align}
where $M$ is the neutron-star mass, and $k_2$ is the second Love number~\citep[see][which includes methods for calculating the second Love number]{hinderer08,flanagan08,damour09}.
The tidal deformability affects the phase of the gravitational-wave signal in neutron star mergers, and is therefore an observable property of the system; we return to this below.

For our grid of candidate equations of state, we have three further conditions before they become \textit{bona fide} candidates. First, we ensure that they support a maximum neutron star mass of at least $2.0\,M_\odot$ to ensure they are conservatively consistent with pulsar mass measurements~\citep{cromartie20,fonseca21}. Second, we ensure causality such that everywhere $c_s\le1$. Third, we enforce thermodynamic consistency such that pressure and energy density are both strictly monotonically increasing. These criteria reduce our number of candidates to 97 $m=$ none equations of state, 327,159 $m=1$, 6,306,578 $m=2$, and 2,042,037 $m=3$ equations of state.

\section{Statistical methodology}
We build up our statistical methodology in two parts. The first develops the method for determing the existence and/or order of a phase transition from a single binary merger---Section~\ref{sec:single_source}. The output of this are both Bayes factors comparing order hypotheses, as well as overall order-level posterior probabilities. We then generalise to infer Bayes factors and order-level posteriors from a population of events in Section~\ref{sec:population_method}

\subsection{Single sources}\label{sec:single_source}
An output of gravitational-wave observations of binary neutron star mergers is marginalised posterior distributions for the tidal deformability and mass: $p(\theta|d)$, where $\theta=\{M_1, M_2, \Lambda_1, \Lambda_2\}$, $M_1\ge M_2$, subscripts denote the primary and secondary components, and $d$ is the gravitational-wave data. (In the case of neutron star-black hole binaries, $\theta$ is the tidal deformability and mass for the individual neutron star involved in the merger.) These posteriors come from standard parameter estimation\footnote{For an introduction to Bayesian inference in gravitational-wave astronomy, see~\citet{thrane19}. In this work, we use the \textsc{Bilby} Bayesian Inference LiBrarY to perform this initial parameter estimation~\citep{ashton19,romeroshaw20}.} that assumes an agnostic prior on the tidal deformabilities $\pi_0(\Lambda_1,\Lambda_2)$. In other words, the priors do not use a specific equation of state. 

For a given equation-of-state hypothesis $\mathcal{H}$ (i.e., a microscopic equation of state plus specific phase-transition parameters), we solve the Tolman-Oppenheimer-Volkoff equations of hydrostatic equilibrium and the perturbation equations to give the relation $\Lambda=\Lambda_{\mathcal H}(M)$. The model prior for $\Lambda_{1,2}$ can then be expressed as the product of two delta functions, one for each of the progenitor components:
\begin{align}
    \pi_\mathcal{H}(\Lambda_1,\Lambda_2|M_1,M_2)=\delta\left(\Lambda_1-\Lambda_\mathcal{H}(M_1)\right)\delta\left(\Lambda_2-\Lambda_\mathcal{H}(M_2)\right).
\end{align}
The evidence for a given hypothesis is the standard marginalised posterior
\begin{align}
    \mathcal{Z}_\mathcal{H}=\int dM_1\int dM_2\pi(M_1,M_2)p\left(d|\theta_{\mathcal{H}}\right),\label{eq:candidate_evidence}
\end{align}
where $\theta_\mathcal{H}=\{M_1,M_2,\Lambda_\mathcal{H}(M_1),\Lambda_\mathcal{H}(M_2)\}$. As $p(d|\theta)\propto p(\theta|d)/\pi_0(\theta)$, we can rewrite this as an importance-sampling integral over the posterior samples
\begin{align}
    \mathcal{Z}_\mathcal{H} \propto \int dM_1\int dM_2\pi(M_1,M_2)\frac{p\left(\theta_\mathcal{H}|d\right)}{\pi_0\left(\theta_\mathcal{H}\right)}.\label{eq:importance_sampling_integral}
\end{align}

In reality, the posteriors from gravitational-wave parameter estimation are represented by a finite number of discrete posterior samples rather than a continuous probability distribution. Following the general importance-sampling strategy of~\citet{lackey15}, we introduce finite-width kernels $K_\sigma$ for $\Lambda$ (we treat mass as exact). With $N$ posterior samples, the posterior distribution can be written as
\begin{align}
    p(\theta|d)\approx\frac{1}{N}\sum_{k=1}^{N}\prod_{j=1}^{2}\delta\left(M_j-M_j^{(k)}\right)\,K_\sigma\left(\Lambda_j-\Lambda_j^{(k)}\right),
\end{align}
where we use $K_{\sigma}(x)=\exp\left(-x^2/2\sigma^2\right)$. 
This can be evaluated at the hypothesis point $\theta_\mathcal{H}$ by substituting $\Lambda_{j}\to\Lambda_{\mathcal H}(M_{j})$ for $j\in\{1,2\}$ such that
\begin{align}
    p(\theta_\mathcal{H}|d)\approx\frac{1}{N}\sum_{k=1}^N\prod_{j=1}^{2}\delta\left(M_j-M_j^{(k)}\right)K_\sigma\left(\Lambda_\mathcal{H}(M_j)-\Lambda_j^{(k)}\right).\label{eq:posterior_thetaH}
\end{align}
The posterior in Equation~\ref{eq:posterior_thetaH} is substituted into Equation~\ref{eq:importance_sampling_integral} such that the evidence becomes
\begin{align}
       \mathcal{Z}_\mathcal{H}&\propto\frac{1}{N}\sum_{k=1}^{N}\frac{\pi\left(M_1^{(k)},M_2^{(k)}\right)}{\pi_0\left(\theta_\mathcal{H}^{(k)}\right)}\prod_{j=1}^{2}K_\sigma\left(\Lambda_\mathcal{H}(M_j^{(k)})-\Lambda_j^{(k)}\right)\notag\\
    &=\frac{1}{N}\sum_{k=1}^{N}\prod_{j=1}^{2}K_\sigma\left(\Lambda_\mathcal{H}(M_j^{(k)})-\Lambda_j^{(k)}\right),\label{eq:evidence_for_one_curve}
\end{align}
and, in the last step, we assume there is no reweighting of the prior on the component masses, and the prior on $\Lambda_\mathcal{H}$ is uniform.

Above, the hypothesis $\mathcal{H}$ refers to a specific candidate equation of state. That is, a single choice of the base hadronic model together with one specific set of phase transition parameters; i.e., the choice of phase-transition order and the associated transition parameters discussed in Section~\ref{sec:phase_transitions}. To obtain the evidence for a phase-transition order $m\in\left\{\rm{none},1,2,3\right\}$ as a category, we marginalise over the choice of base hadronic equation of state and specific phase-transition parameters. Throughout, the candidates entering this sum are restricted to those built from the shared quark-matter branch described in Section~\ref{sec:phase_transitions} (except for $m={\rm none}$, which are simply the unmodified baseline equations of state), so that base hadronic equation of state and transition order are the only two things varying between candidates of different $m$. Implicitly assuming a uniform prior over the grid of candidates, this becomes
\begin{align}
    \mathcal{Z}_m=\frac{1}{N_m}\sum_{i\in m}\mathcal{Z}_{\mathcal{H}_i},\label{eq:m_evidence}
\end{align}
where $N_m$ is the number of candidates of phase-transition order $m$ in our grid of base equations of state and transition parameters. Taking the average in Equation~\ref{eq:m_evidence} implicitly applies an Occam penalty; that is, phase-transition orders with more free parameters are statistically penalised. 
The order-level posterior probability, assuming equal prior odds between orders, is
\begin{align}
    p(m|d)=\frac{\mathcal{Z}_m}{\displaystyle\sum_{m'}\mathcal{Z}_{m'}},\label{eq:single_posterior}
\end{align}
where, explicitly, the denominator sums over the total evidence of all phase-transition orders $m'\in\{{\rm none}, 1, 2, 3\}$ to ensure normalisation of each $p(m|d)$.

Finally, we calculate the Bayes factor between two competing transition orders---e.g., between a first-order phase transition $m_1$ and second-order phase transition $m_2$---as
\begin{align}
    \mathcal{B}_{m_i/m_j}=\frac{\mathcal{Z}_{m_i}}{\mathcal{Z}_{m_j}}.\label{eq:single_Bayes}
\end{align}
We again note that, in the absence of a principled alternative, we assign the prior odds between any two phase orders a value of one
\begin{align}
    \mathcal{O}_{m_i/m_j}=1\quad\forall\quad m_i,m_j\in\left\{{\rm none},1,2,3\right\},
\end{align}
such that the Bayes factor and the odds are equal.

In addition to calculating evidence for the transition order, we infer the parameters of the transition itself. Let $\xi$ denote one transition parameter, for example the transition density $n_h$, and let $N_{m, \xi}$ be the number of candidates of phase-transition order $m$ with that value of $\xi$. The marginal posterior of $\xi$ for a given $m$ is
\begin{align}
    p(\xi|d,m) = \frac{1}{N_m\,\mathcal{Z}_m}\sum_{i\in m}\delta_{\xi_i\xi}\,\mathcal{Z}_{\mathcal{H}_i},\label{eq:marginal_single}
\end{align}
where $\xi_i$ is the value of $\xi$ for candidate $\mathcal{H}_i$, and $\delta_{\xi_i\xi}$ is the Kronecker delta. This implicitly uses the same prior as above, such that the marginal prior is
\begin{align}
    \pi(\xi|m)=\frac{N_{m,\xi}}{N_m}.\label{eq:marginal_prior}
\end{align}

Finally, if we simply assume the hypothesis that there is a phase transition $T$, then the marginalised posterior for $\xi$ becomes the evidence-weighted sum of the posteriors
\begin{align}
    p(\xi|d,T) = \sum_{m\in\mathcal{S}_\xi}\frac{\mathcal{Z}_m}{\displaystyle\sum_{m'\in\mathcal{S}_\xi}\mathcal{Z}_{m'}}\,p(\xi|d,m),\label{eq:marginal_single_mixture}
\end{align}
with corresponding prior
\begin{align}
    \pi(\xi|T) = \frac{1}{|\mathcal{S}_\xi|}\sum_{m\in\mathcal{S}_\xi}\pi(\xi|m),\label{eq:marginal_mixture_prior}
\end{align}
where $\mathcal{S}_\xi\subseteq\{1,2,3\}$ is the set of transition orders for which $\xi$ is defined; e.g., $\mathcal{S}_{n_h}=\{1,2,3\}$ and $\mathcal{S}_{n_2}=\{2,3\}$. We note that these effective priors are not necessarily flat in each parameter because some combinations of parameters presented in Table~\ref{tab:transition_grids} result in neutron stars that do not satisfy our quality-control criteria (e.g., not supporting neutron stars with $2\,M_\odot$). We return to this in later sections.

\subsection{Population}\label{sec:population_method}
From the previous sections, for event $i$ and candidate equation-of-state curve $E_j$ we have the likelihood of that event given the data $\mathcal{L}_i\left(d_i|E_j\right)$. In this language, the evidence for an $m^{\rm th}$-order phase transition from event $i$ is
\begin{align}
    {\mathcal Z}_{m,i} = \frac{1}{|\mathcal{E}_{m}|}\sum_{j \in \mathcal{E}_{m}} p(d_i|E_j),
\end{align}
where $\mathcal{E}_{m}$ is the set of candidate equations of state of order $m$.  We note this uses the mean, not the sum (i.e., in the above equation we divide by $|\mathcal{E}_m|$), because we cannot guarantee the same number of equation-of-state candidates for each phase-transition order. 

The population-level evidence for the $m^{\rm th}$-order phase transition ${\mathcal Z}^{\rm pop}_m$ multiplies the likelihood over all events and then marginalises over the candidate grid~\citep{mandel19,landry19}
\begin{align}
    {\mathcal Z}^{\rm pop}_m = \frac{1}{|\mathcal{E}_{m}|}\sum_{j \in \mathcal{E}_{m}}\prod_{i=1}^{N} p(d_i|E_j).
\end{align}
The population posterior probability for any given order is 
\begin{align}
    p(m|\{d_i\}) = \frac{\mathcal{Z}^{\rm pop}_m}{\displaystyle\sum_{m'}\mathcal{Z}^{\rm pop}_{m'}},\label{eq:pop_prob}
\end{align}
where again the sum over $m'\in\{{\rm none}, 1, 2, 3\}$ in the denominator to assure normalised probabilities, and we implicitly assume a flat prior between all $m$ orders (i.e., $\pi(m)=1/4$).

Again assuming equal odds between phase-transition-order hypotheses, the Bayes factor between two competing transition orders is
\begin{align}
    B_{m_i/m_j}=\frac{\mathcal{Z}^{\rm pop}_{m_i}}{\mathcal{Z}^{\rm pop}_{m_j}}.\label{eq:pop_Bayes}
\end{align}

The population generalisation of the marginalised posteriors (Equation~\ref{eq:marginal_single}) becomes
\begin{align}
    p(\xi|\{d_i\},m) = \frac{1}{|\mathcal{E}_m|\,\mathcal{Z}^{\rm pop}_m}\sum_{j\in\mathcal{E}_m}\delta_{\xi_j\xi}\prod_{i=1}^{N}p(d_i|E_j),\label{eq:marginal_pop}
\end{align}
and the prior is unchanged from Equations~\ref{eq:marginal_prior} and~\ref{eq:marginal_mixture_prior}. Again, conditioned on the existence of a transition, the evidence-weighted marginalised population posterior is 
\begin{align}
    p(\xi|\{d_i\},T) = \sum_{m\in\mathcal{S}_\xi}\frac{\mathcal{Z}^{\rm pop}_m}{\displaystyle\sum_{m'\in\mathcal{S}_\xi}\mathcal{Z}^{\rm pop}_{m'}}\,p(\xi|\{d_i\},m).\label{eq:marginal_pop_mixture}
\end{align}

\section{Single-sources}
\subsection{GW170817}\label{sec:170817}
To date there have been two gravitational-wave observations of binary neutron star mergers, and a couple neutron star-black hole mergers. By far, the most constraining of these in terms of tidal deformability is the spectacular binary neutron star merger GW170817~\citep{abbott17_GW170817detection,abbott17_GW170817MMA} that was observed in gravitational waves and across the electromagnetic spectrum. We do not expect GW170817 to be constraining in terms of phase transitions, but we use this as a nice test case for our methodology before moving to higher signal-to-noise ratio synthetic mergers and populations of mergers.

We use posterior samples for GW170817 from the LIGO/Virgo GWTC-1 parameter estimation data release\footnote{posterior samples can be found at \doi{10.7935/KSX7-QQ51}.}, specifically using the \texttt{IMRPhenomPv2NRTidal} waveform~\citep{dietrich17} with the low-spin prior~\citep{abbott18_GW170817nsradii,abbott19_gwtc1}.

\begin{figure*}
    \centering
    \includegraphics[width=0.98\textwidth]{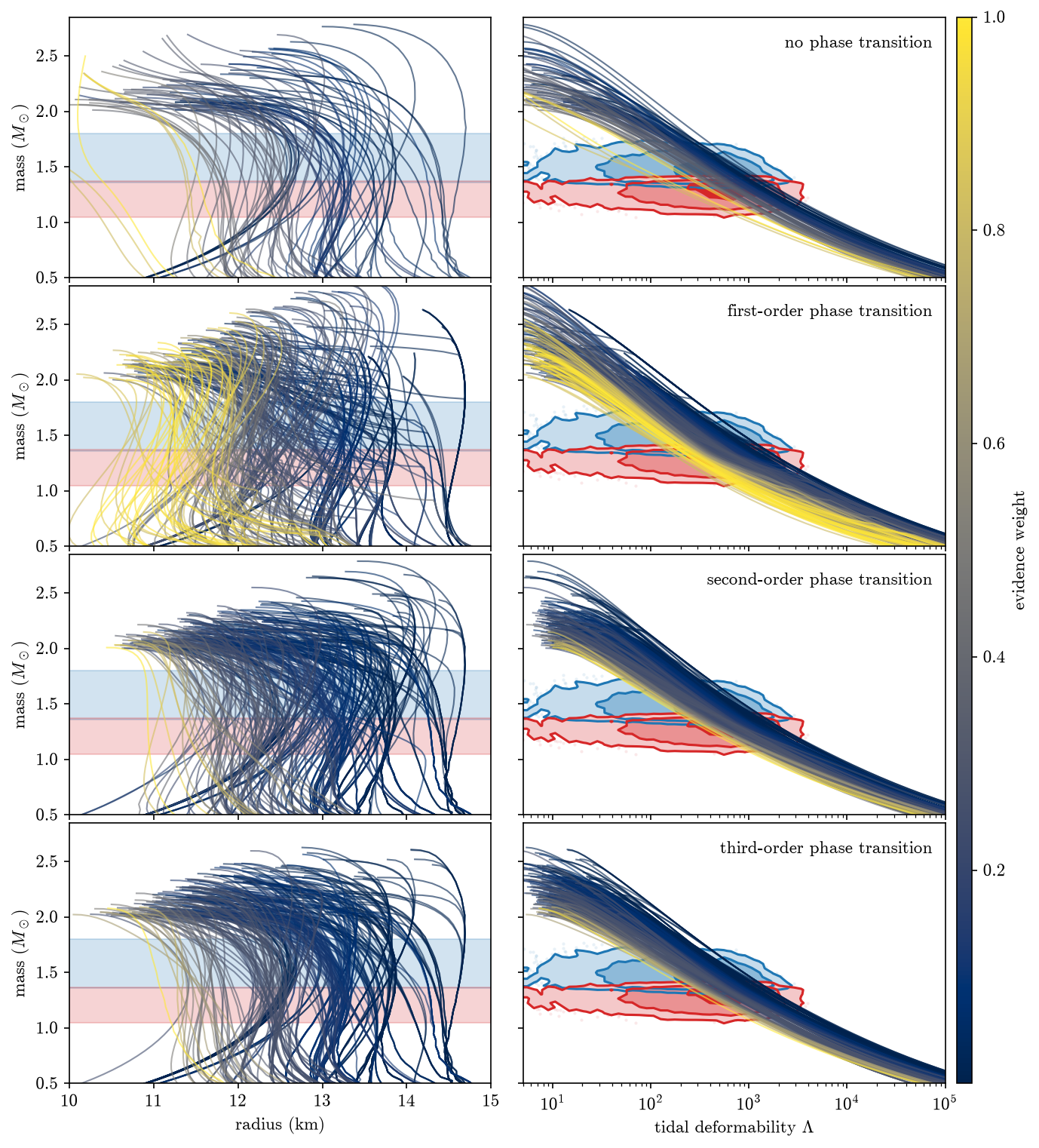}
    \caption{Normalised evidence for each candidate equation of state for GW170817. The blue and red contours are the posterior distributions for the gravitational-wave data. Each row shows different phase-transition order as described in the label of each row, and curve normalisation for the evidence weight (i.e., colour of the curve) is done per row. To aid in visualisation, in the bottom three rows, we downsample to 300 equation-of-state curves. }
    \label{fig:170817_massradiuslambda}
\end{figure*}

Figure~\ref{fig:170817_massradiuslambda} is a mess of mass-radius (left column) and mass-tidal deformability (right column) curves. We colour each curve by its evidence $\mathcal{Z}_\mathcal{H}$ (Equation~\ref{eq:evidence_for_one_curve}) normalised to the most favoured curve in each row, i.e., the same phase-transition order. The colours therefore cannot be compared between rows.

The data clearly distinguish individual curves, showing they prefer (more yellow colour) soft equations of state (i.e., with smaller radii at fixed mass) than stiff equations of state (i.e., equations of state with larger radii at fixed mass are more blue). This is consistent with other analyses of GW170817 that indicate soft equations of state are preferred such that radii at $1.4\,M_\odot$ are $\lesssim\unit[13]{km}$~\citep[e.g.,][]{abbott18_GW170817nsradii,annala18,most18,fattoyev18}.

We calculate the individual order-level posterior probability $p(m|d)$ from Equation~\ref{eq:single_posterior} as well as the Bayes factors between all competing transition orders $\mathcal{B}_{m_i/m_j}$ from Equation~\ref{eq:single_Bayes}. These are shown in Table~\ref{tab:bayes_factors_gw170817}, which shows tantalising structure that a first-order transition may be preferred over a purely hadronic star or a second- or third-order phase transition. 

While the evidence above is certainly not statistically significant, it is interesting! It is also worth noting that this is consistent with other similar studies in the literature that suggest GW170817 posteriors are consistent with phase transitions in general~\citep{essick20, blaschke20,annala20}, and even consistent with first-order phase transitions~\citep{cao26}\footnote{\citet{cao26} and the results presented herein are not directly comparable. In the former paper, their phase transition density is most likely above the neutron star central density of the most massive neutron stars, and is therefore attributable to the reconciling pQCD calculations with heavy pulsars, rather than directly from inference of gravitational-wave data.} On the other hand, there are also works that find no significant evidence for phase transitions~\citep{pang20,essick23}, further suggesting we should not read too much into our inconclusive statistics.

\begin{table}
    \centering
    \caption{Order-level posterior probabilities $p(m|d)$ and pairwise Bayes factors $\mathcal{B}_{m_i/m_j}$---see Equation~\ref{eq:single_Bayes}---between phase-transition order classes $m_i,m_j \in \{\rm none,1,2,3\}$ for the GW170817 posterior. The convention we use is $m_i$ are rows and $m_j$ are the columns.}
    \label{tab:bayes_factors_gw170817}
    \begin{tabular}{| l | cccc|}
        \hline
        & {none} & {first} & {second} & {third} \\
        \hline
        $p(m|d)$ & 0.23 & 0.41 & 0.17 & 0.19\\
        \hline
        \hline
        none   & & 0.56 & 1.33 & 1.23 \\
        first  & & & 2.37 & 2.20 \\
        second & & & & 0.93 \\
        \hline
    \end{tabular}
\end{table}

\begin{figure*}
    \centering
    \includegraphics[width=0.98\textwidth]{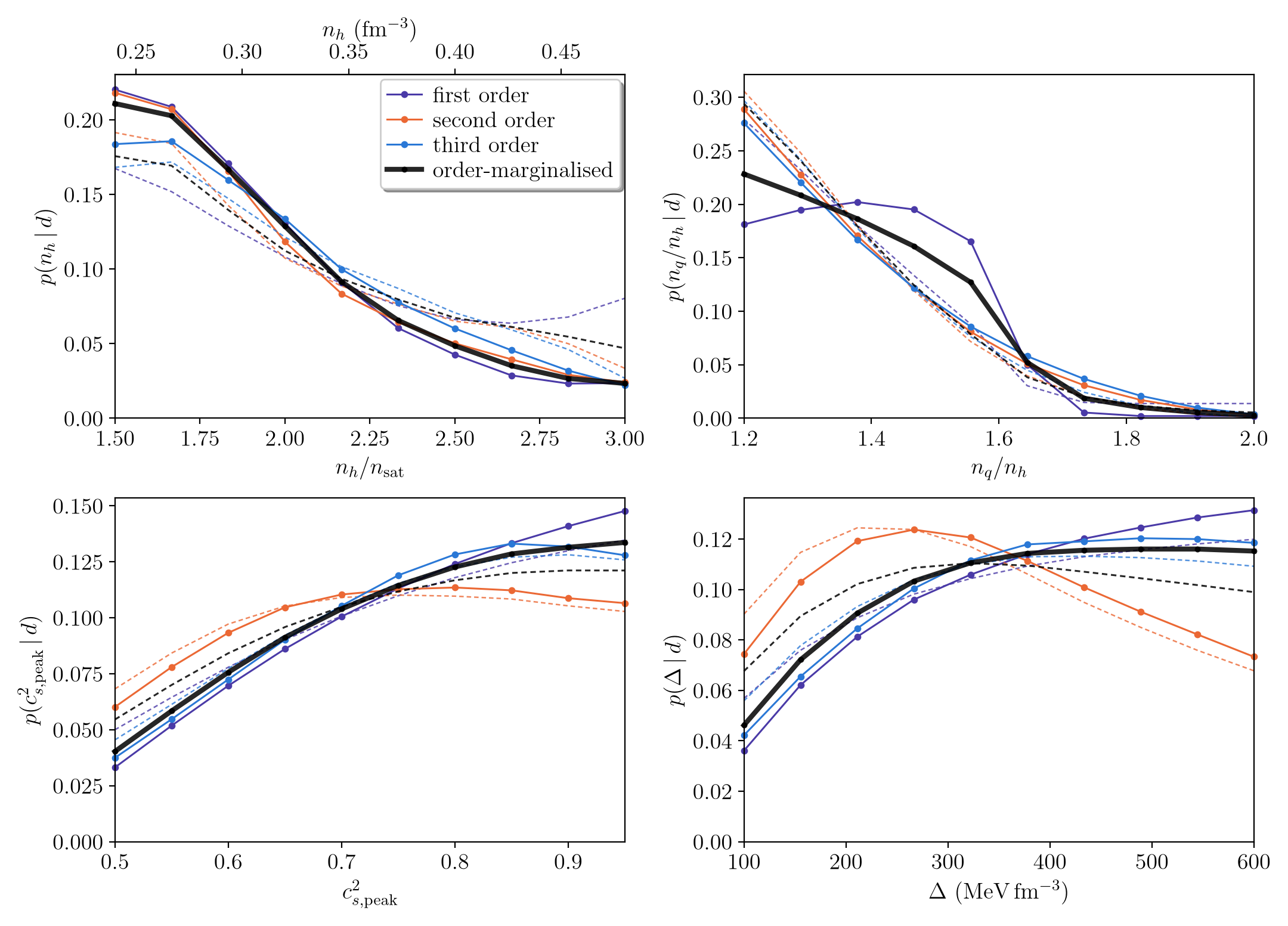}
    \caption{Marginalised posterior distributions for GW170817 of the transition onset density $n_h$ (top left), density jump $n_q/n_h$ (top right), peak sound speed $c_{s,\rm peak}^2$ (bottom left) and relaxation width $\Delta$ (bottom right). We show posteriors assuming first- (purple), second- (orange) and third-order (blue) phase transitions, as well as the order-transition marginalised curve (black). Solid curves are posteriors, and dashed curves are priors.}
    \label{fig:170817_marginals}
\end{figure*}

Figure~\ref{fig:170817_marginals} shows the one-dimensional marginal posterior distributions for a subset of the phase-transition parameters, specifically $n_h$ (top left), $n_q/n_h$ (top right)  $c_s^2$ (bottom left), and $\Delta$ (bottom right). In each figure, we show both the posterior (solid curves) and effective prior (dashed curves) for each of the first-, second-, and third-order phase transitions, as well as the order-transition marginalised posterior (Equation~\ref{eq:marginal_single_mixture}). We do not show other marginal posteriors as they are largely uninformative, and entirely uninteresting.

It is tempting to see structure in these plots. However, we caution the reader that the posterior distributions are not all that distinct from the prior distributions in each case (other than some slight structure in $n_q/n_h$), which takes their shapes because of the quality-control cuts we make on equations of state curves that e.g., do not meet the minimal $2\,M_\odot$ threshold, obey causality, or obey monotonicity. The $c^2_{s,\rm peak}$ and $\Delta$ posteriors closely follow their priors, whose shapes are set by the cuts. The $n_h$ posterior shows a mild preference for low onset densities, consistent with the softening discussed above, but remains broad. We also note there is a covariance between $n_h$ and $n_q/n_h$, which we explore further when examining the population results.

\subsection{GW190425}\label{sec:190425}
We `rinse and repeat' the calculations in the above section for the gravitational-wave event GW190425~\citep{abbott20_GW190425}, which was a binary neutron star merger at $\unit[159^{+69}_{-72}]{Mpc}$ (i.e., about four times further away than GW170817) with $M_1=1.75^{+0.17}_{-0.09}\,M_\odot$, $M_2=1.56^{+0.08}_{-0.13}\,M_\odot$ (all 90\% CIs), $\Lambda_1\le 1090$, and $\Lambda_2\le 1662$ (90\% upper limits as they are consistent with $\Lambda_{1,2}=0$).  The results are even less informative than the relatively uninformative GW170817 results. We therefore do not waste paper space by showing another full-page plot, but instead simply quote the $p(m|d)$ and Bayes factor outcomes in Table~\ref{tab:bayes_factors_gw190425} showing that there is nothing interesting to learn here (we remind the reader that the prior has each order at $\pi(m)=0.25$). The marginal posterior distributions are equally uninteresting, and are therefore not shown to maintain some sense of brevity.

\begin{table}
    \centering
    \caption{Same as Table~\ref{tab:bayes_factors_gw170817}, but for GW190425. Note, the extra decimal places are required as the $p(m|d)$ values are otherwise close to indistinguishable from their prior value of 0.25, and the Bayes factors are nearly indistinguishable from unity.}
    \label{tab:bayes_factors_gw190425}
    \begin{tabular}{| l | cccc|}
        \hline
        & {none} & {first} & {second} & {third} \\
        \hline
        $p(m|d)$ & 0.2539 & 0.2482 & 0.2460 & 0.2519\\
        \hline
        \hline
        none   & & 1.0230 & 1.0321 & 1.0079 \\
        first  & & & 1.0089 & 0.9853 \\
        second & & & & 0.9766 \\
        \hline
    \end{tabular}
\end{table}

\subsection{Synthetic mergers in A$^+$ and  A$^\sharp$}
We create GW170817-like binary neutron star mergers and inject them into noise with the A$^+$~\citep{fritschel25} and A$^\sharp$~\citep{sun26} sensitivity curves. We choose masses such that the chirp mass and mass ratio are consistent with GW170817, and a luminosity distance of $d_L=\unit[40]{Mpc}$. We draw random values of the tidal deformabilities from the standard prior such that our injection has $m_1=1.38\,M_\odot$, $m_2=1.34\,M_\odot$, $\Lambda_1=483$, and $\Lambda_2=771$. We emphasise that these are not associated with any known underlying equation of state from our set, but instead are chosen simply to match the posterior distributions from the observed GW170817.  
We use only the two LIGO (Hanford and Livingston) observatories at those sensitivities, which yields network SNRs of 76 and 95, respectively.\footnote{These SNRs are lower than the values quoted in the Introduction because, while our injection shares GW170817's masses, tidal deformabilities and distance, its orientation and sky location differ. The orientation alone reduces the SNR by a factor of $\approx1.9$.} 

To speed up the otherwise prohibitively-expensive inference calculations, we calculate reduced-order models for the waveforms and evaluate the likelihood using the reduced-order quadrature (ROQ) framework~\citep{smith16}. We construct our ROQs using the \texttt{gwRombusX} package~\citep{guttman26}, an extension of the \texttt{Rombus}\footnote{\url{https://github.com/ADACS-Australia/rombus}} reduced-order modelling framework and its gravitational-wave extension \texttt{gwRombus}~\citep{baker25}. \texttt{gwRombusX} is designed to support long-duration signals through disk-backed storage of large bases, streaming linear algebra, adaptive frequency sampling, and frequency subbanding, which therefore allows significantly faster calculations in both A$^+$/A$^\#$ era and next-generation observatories such as Cosmic Explorer and Einstein Telescope---see the next Section for details. 

\begin{figure*}
    \centering
    \includegraphics[width=0.98\textwidth]{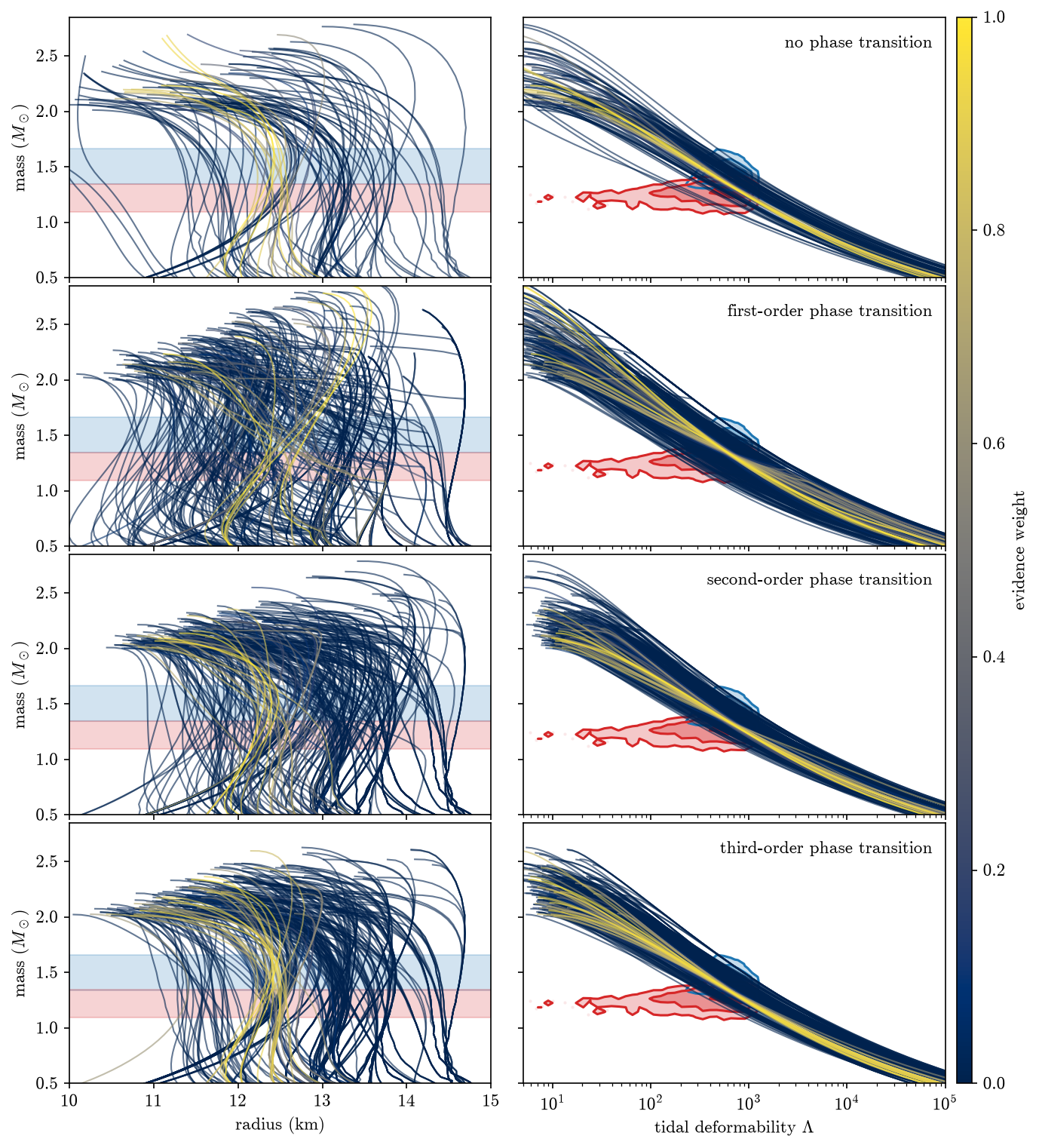}
    \caption{Same as Figure~\ref{fig:170817_massradiuslambda}, but for the GW170817-like synthetic injection into two A$^\#$ LIGO observatories.}
    \label{fig:170817_Asharp_massradiuslambda}
\end{figure*}

Figure~\ref{fig:170817_Asharp_massradiuslambda} shows the normalised evidences for our candidate equations of state curves for our A$^\sharp$ injection (we do not show the A$^+$ injection as these are big plots, but instead simply report the final numbers below). The mass and tidal deformability posterior distributions from parameter estimation are seen to be considerably tighter (especially $\Lambda_1$, the blue posterior in the right-hand column) than for the real GW170817 observation due simply to the significantly-improved detector sensitivity. One can see the impact this has on the evidence weights for the respective equation-of-state curves. 

The change in evidences for the individual equation-of-state curves translates into more interesting mode-level Bayes factors and posterior values, which are shown for both the A$^+$ and A$^\#$ simulations in Table~\ref{tab:bayes_factors_gw170817_Aplus_Asharp}. It is important to stress that the underlying equations of state used in these injections are not real, and therefore any structure in these numbers is purely coincidental. We remedy the false nature of the injection in Section~\ref{sec:synthetic_populations} when looking at populations of binary mergers, and also save detailed discussion of marginal distributions to that section as well.

\begin{table}
    \centering
    \caption{Same as Table~\ref{tab:bayes_factors_gw170817}, but for the GW170817-like synthetic injections into A$^+$ and A$^\#$. }
    \label{tab:bayes_factors_gw170817_Aplus_Asharp}
    \begin{tabular}{| l | cccc|}
        \hline
        & {none} & {first} & {second} & {third} \\
        \hline
        & \multicolumn{4}{c|}{A$^+$} \\
        \hline
        $p(m|d)$ & 0.30 & 0.21 & 0.24 & 0.25\\
        \hline
        \hline
        none   & & 1.42 & 1.27 & 1.22 \\
        first  & & & 0.89 & 0.86 \\
        second & & & & 0.96\\
        \hline
        \hline
        & \multicolumn{4}{c|}{A$^\#$} \\
        \hline
        $p(m|d)$ & 0.31 & 0.21 & 0.23 & 0.25\\
        \hline
        \hline
        none   & & 1.48 & 1.35 & 1.25 \\
        first  & & & 0.92 & 0.84 \\
        second & & & & 0.92 \\
        \hline
    \end{tabular}
\end{table}

\begin{figure*}
    \centering
    \includegraphics[width=0.98\textwidth]{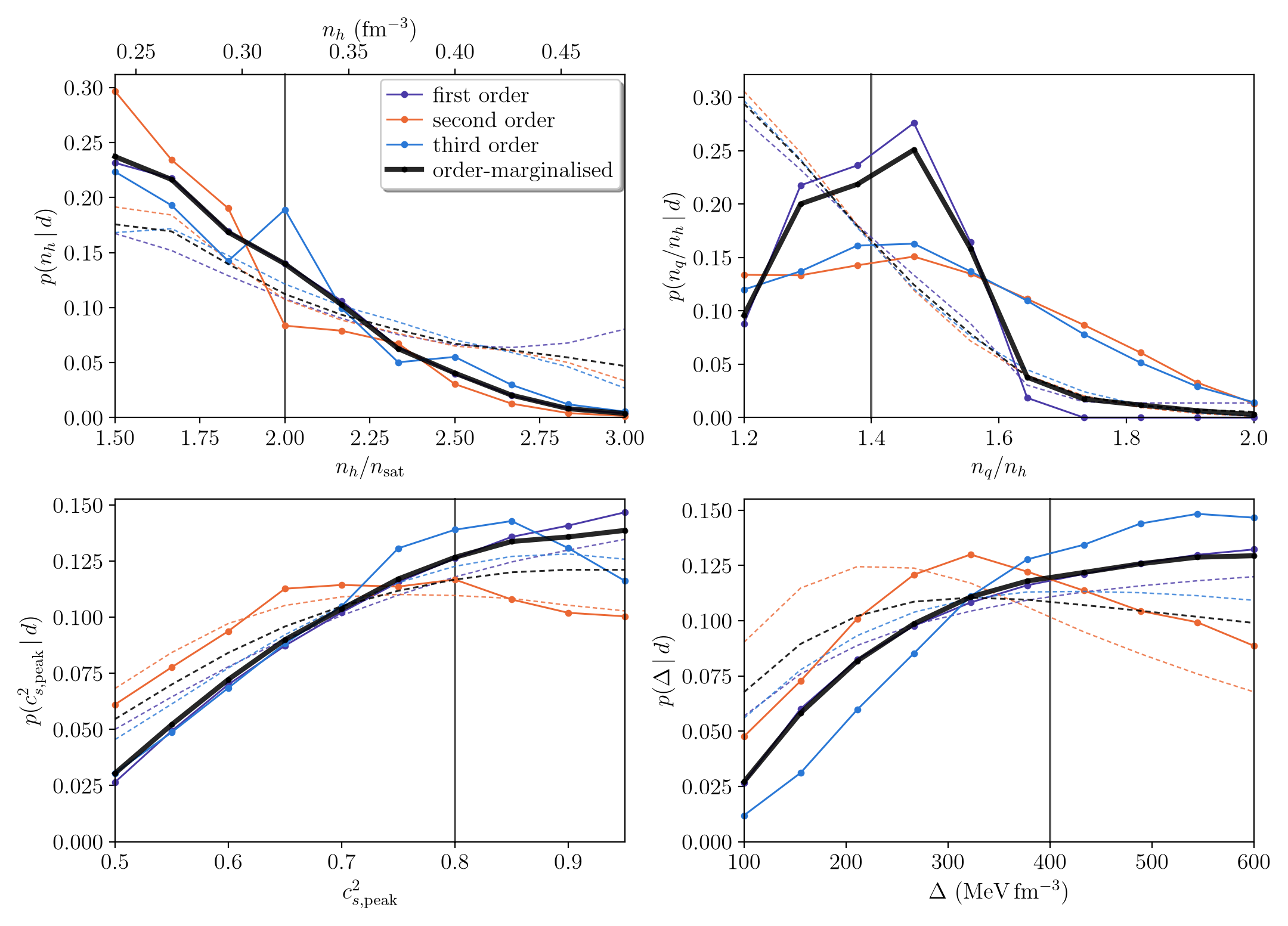}
    \caption{Same as Figure~\ref{fig:170817_marginals}, but for the Cosmic Explorer injection. Vertical grey lines mark the injected values: $n_h=2\,n_{\rm sat}$, $n_q/n_h=1.4$, $c_{s,\rm peak}^2=0.8$ and $\Delta=\unit[400]{MeV\,fm^{-3}}$.}
    \label{fig:ce_marginals}
\end{figure*}

Figure~\ref{fig:ce_marginals} shows the one-dimensional marginal distributions for this injection (the same subset of parameters as Figure~\ref{fig:170817_marginals}). One can see here that $n_q/n_h$ has a posterior peak at the correct value, albeit with a broad recovery, while the other three parameters are largely uninformative, and remain consistent with their priors.

\subsection{Synthetic merger in the XG era}
We create a single synthetic binary neutron star merger, and inject it into a single Cosmic Explorer observatory~\citep{abbott17_CE}. Our injections in the previous section resulted in relatively broad posterior distributions (compared to this section), and so it did not matter that our mass and tidal-deformability values were not drawn from a specific equation of state in our set. However, for Cosmic Explorer, the posteriors are significantly more constrained, implying a poor choice could lead to spurious results.  

We somewhat arbitrarily choose the baseline RG(SLY9) equation of state, and impose a first-order phase transition with $n_h=2.0\,n_{\rm sat}$, $n_q/n_h=1.4$, $c_{s,{\rm peak}}^2=0.8$, and $\Delta=\unit[400]{MeV\,fm^{-3}}$. We choose source-frame component masses $M_1=1.3450\,M_\odot$ and $M_2=1.3079\,M_\odot$, which have corresponding values of $\Lambda_1=296$ and $\Lambda_2=357$. We inject the signal at $d_L=\unit[100]{Mpc}$.

We analyse the full signal between 5 and $\unit[2048]{Hz}$, which requires a full $\unit[7,500]{s}$ segment duration and $\unit[4096]{Hz}$ sampling. This necessarily requires inclusion of Earth's rotation into the detector antenna-pattern functions, as well as full consideration of the free-spectral range. As such, we closely follow the work of~\citet{guttman26}, including the construction of our ROQ using \texttt{gwRombusX} and the full thirteen subbands.

This configuration has an optimal SNR=418 which, combined with the long segment duration and high sampling frequency, makes it a computationally challenging calculation. We use \texttt{pBilby} with the \texttt{dynesty} sampler and 4,000 live points across 256 cores; the sampling takes $\approx20$ hours of walltime.

\begin{figure*}
    \centering
    \includegraphics[width=0.98\textwidth]{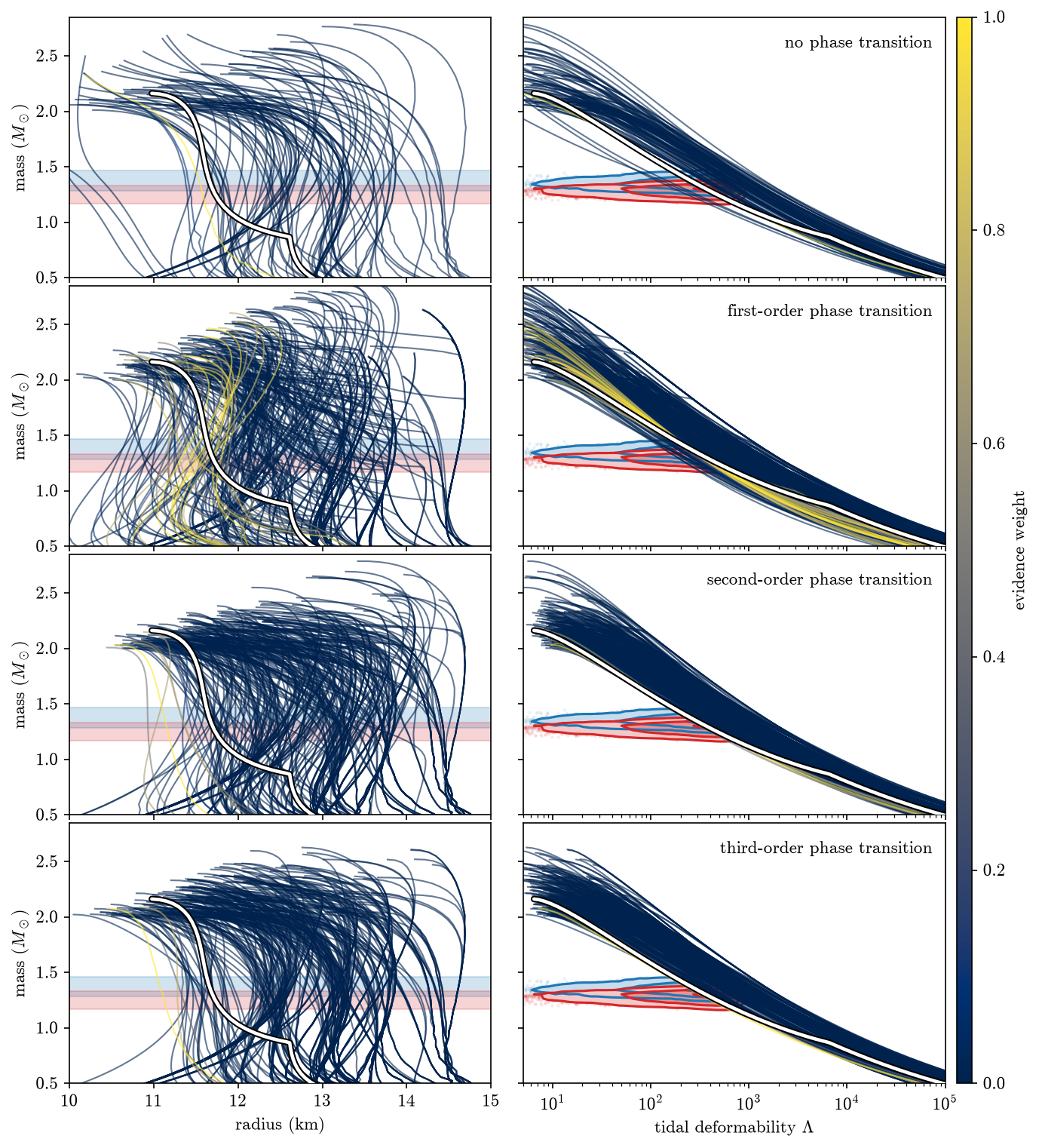}
    \caption{Same as Figures~\ref{fig:170817_massradiuslambda} and \ref{fig:170817_Asharp_massradiuslambda}, but for our Cosmic Explorer injection at 100\,Mpc. The white curve with black outline is the equation of state used for the injection.}
    \label{fig:CE_EOSs}
\end{figure*}

Our full Bayesian analysis leads to posteriors for the masses and tidal deformabilities that are significantly more constrained than those for our A$^\sharp$ synthetic merger, despite the 2.5 times larger distance. In particular, the 90\% confidence-interval widths for the masses reduces from 0.19 and $0.16\,M_\odot$ to 0.08 and $0.08\,M_\odot$ for the primary and secondary masses respectively, and 590 and 830 to 527 and 616 for the primary and secondary tidal deformabilities, respectively. Figure~\ref{fig:CE_EOSs} shows the weighted evidences for the equations of state for our Cosmic Explorer synthetic injection, akin to Figures~\ref{fig:170817_massradiuslambda} and Figures~\ref{fig:170817_Asharp_massradiuslambda}. We note the posterior distributions for $\Lambda$ in the right-hand column appear broader than for A$^\sharp$ in Figure~\ref{fig:170817_Asharp_massradiuslambda}, but note this is due to 1) the logarithmic horizontal scale and 2) the fact that we measure the combination $\tilde{\Lambda}$ better than individual $\Lambda$'s, and this is a nearly equal-mass binary, implying $\Lambda_1$ and $\Lambda_2$ are strongly anti-correlated.

Figure~\ref{fig:CE_EOSs} shows that the tightly-constrained mass and tidal-deformability posterior distributions corresponds to significantly fewer equation of state curves being strongly preferred over others. The corresponding order-level posterior probabilities and Bayes factors are shown in Table~\ref{tab:bayes_factors_ce}. Here we see that the first-order phase transition is significantly favoured ($p(m=1|d)=0.74$) over the other orders, with corresponding Bayes factors $\mathcal{B}_{1 / m}=12.8$, $7.8$, and $7.2$ for $m\in\{{\rm none},\,2,\,3\}$.

\begin{table}
    \centering
    \caption{Same as Table~\ref{tab:bayes_factors_gw170817}, but for the Cosmic Explorer injection at $d_L=\unit[100]{Mpc}$. The true equation of state contains a first-order phase transition, which is the preferred model both for the probabilities (top row) and the Bayes factors (third row).}
    \label{tab:bayes_factors_ce}
    \begin{tabular}{| l | cccc|}
        \hline
        & {none} & {first} & {second} & {third} \\
        \hline
        $p(m|d)$ & 0.06 & 0.74 & 0.10 & 0.10\\
        \hline
        \hline
        none   & & 0.08 & 0.61 & 0.56 \\
        first  & & & 7.77 & 7.20 \\
        second & & & & 0.93 \\
        \hline
    \end{tabular}
\end{table}

\section{Populations}
\subsection{GW170817+GW190425}
In an exercise of futility, we calculate and report the population analysis of \textit{all} binary neutron star mergers observed by the LIGO-Virgo-KAGRA collaboration to date. Using the statistical infrastructure built in Section~\ref{sec:population_method}, and the single-source calculations of GW170817 and GW190425 discussed in Sections~\ref{sec:170817} and~\ref{sec:190425}, respectively, we calculate $p(m|\{d\})$, where $\{d\}=\{d_{\rm GW170817},\,d_{\rm GW190425}\}$. These are unsurprisingly unremarkable and uninformative, and shown in Table~\ref{tab:bayes_factors_combined}, where we again remind the reader that the prior is $\pi(m)=0.25$ for all $m$. 

We also calculate the marginal distributions for the combination of these two events, however they are as unremarkable as one would expect, and therefore do not show them.

\begin{table}
    \centering
    \caption{Same as Table~\ref{tab:bayes_factors_gw170817}, but for the combined analysis of GW170817 and GW190425, i.e. $\{d\}=\{d_{\rm GW170817},\,d_{\rm GW190425}\}$. The joint evidence for each order is the mean over candidates of the product $p(d_{\rm GW170817}|\mathcal{H})\,p(d_{\rm GW190425}|\mathcal{H})$, not the product of the per-event means.}
    \label{tab:bayes_factors_combined}
    \begin{tabular}{| l | cccc|}
        \hline
        & {none} & {first} & {second} & {third} \\
        \hline
        $p(m|d)$ & 0.23 & 0.38 & 0.19 & 0.20\\
        \hline
        \hline
        none   & & 0.61 & 1.25 & 1.17 \\
        first  & & & 2.06 & 1.93 \\
        second & & & & 0.93 \\
        \hline
    \end{tabular}
\end{table}

It is worth noting that we could easily add observed neutron star-black hole mergers to this calculation. However, these are equally uninformative, and would not result in any substantive or interesting change to our conclusions.

\subsection{Synthetic populations}\label{sec:synthetic_populations}
We create universes of observations, with each universe having a single equation of state that we consider as the ground truth. We choose a modestly stiff equation of state for our base hadronic model. To make this feel scientific, we calculate the radii of all our hadronic equations of state at $1.4\,M_\odot$, and choose the equation of state with radius closest to $\unit[12.5]{km}$. The lucky winner of this relatively arbitrary process is the zero-temperature and beta-equilibrium unified equation of state RG(SLY9)~\citep{danielewicz09,gulminelli15}. For our other universes, we use the same underlying hadronic equation of state, but introduce first- and second--order phase transitions. For first-order, we use $n_h=2\,n_{\rm sat}$, $n_q/n_h=1.4$, $c^2_{s,\rm peak}=0.8$, $\Delta=\unit[400]{MeV}$. For second-order, we use those same parameters and $n_2=0.85,{\rm fm^{-3}}$ and $\chi_2^{B,h}=\chi_2^{B,q}=\unit[0.001]{fm^{-3}\,MeV^{-1}}$.

We draw neutron star masses from a double-peaked Gaussian, where the lower peak accounts for the observed Galactic double neutron star systems and the upper peak accounts for the masses of the second binary neutron star merger observed in gravitational waves GW190425~\citep{abbott20_GW190425}. This mass distribution is given by
\begin{align}
    p(m)=(1-\epsilon)\mathcal{N}\left(\mu_1,\sigma_1\right)+\epsilon\mathcal{N}\left(\mu_2,\sigma_2\right),
\end{align}
where $\mathcal{N}(\mu,\sigma)$ is a Gaussian of mean $\mu$ and standard deviation $\sigma$. We use $\mu_1=1.32\,M_\odot$, $\sigma_1=0.11\,M_\odot$ consistent with Galactic double neutron stars~\citep[e.g.,][]{farrow19}, $\mu_2=1.80\,M_\odot$, $\sigma_2=0.21\,M_\odot$ consistent with GW190425, and $\epsilon=0.35$ because we have no other idea what to choose~\citep{sarin20}.

For each binary neutron star system, we draw two masses from the above distribution and calculate corresponding $\Lambda$ values for both stars from the equation of states described above. We draw systems uniform in comoving volume, and from all other standard priors (e.g., on sky location, etc). We create waveforms using the \textsc{IMRPhenomPv2\_NRTidal} waveform approximant. 

\begin{figure*}
    \centering
    \includegraphics[width=0.98\textwidth]{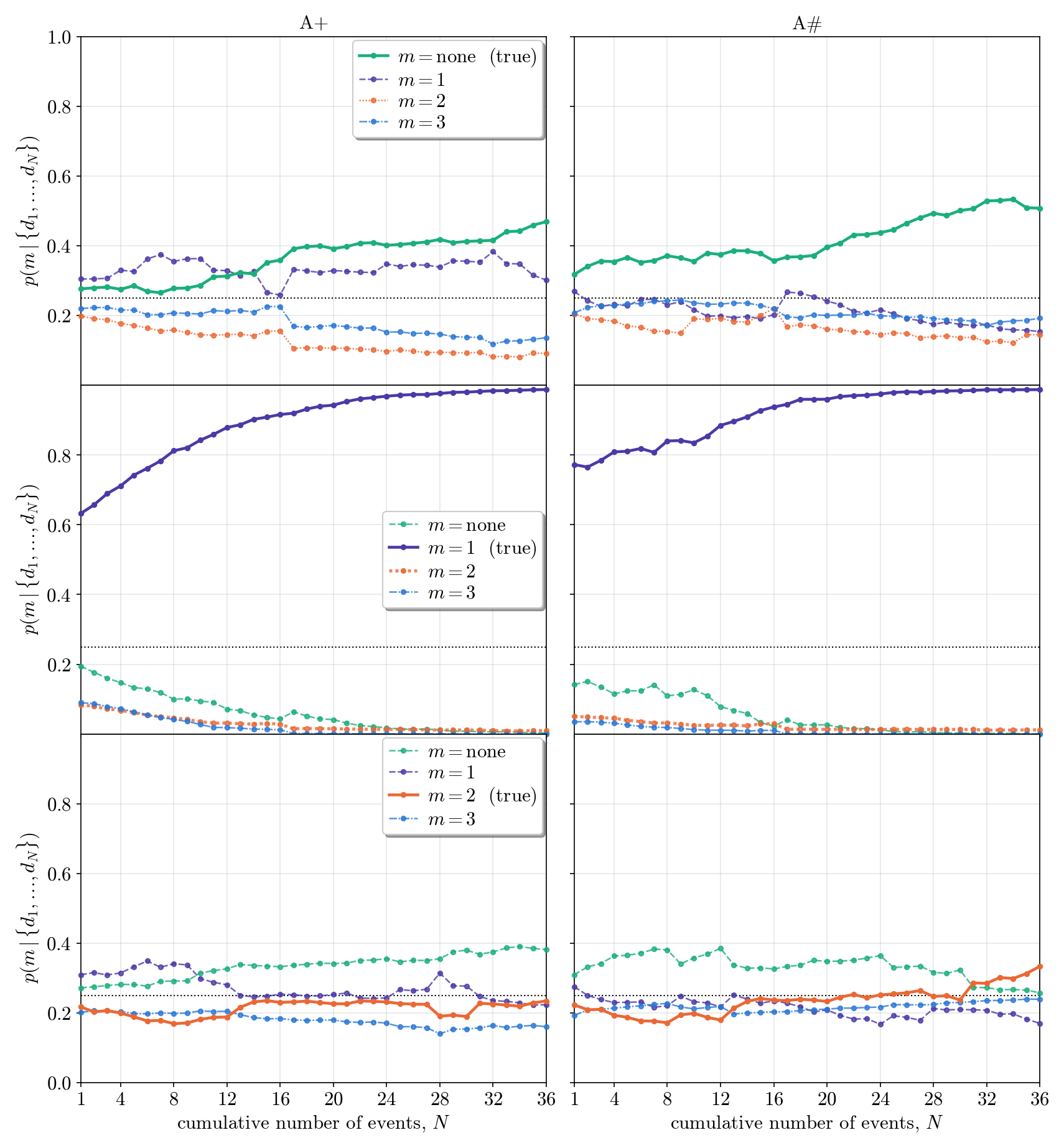}
    \caption{Cumulative order-level posterior distributions. The left and right columns are injections into A+ and A$^\sharp$ networks, respectively. The top, middle, and bottom rows show the injections with true equations of state being no phase transition, first-order phase transition and second-order phase transition, respectively.}
    \label{fig:population_probability}
\end{figure*}

Figure~\ref{fig:population_probability} shows the cumulative order-level posterior distribution as a function of the number of events. The left column shows synthetic injections into A+, and the right column into A$^\sharp$. The top, middle, and bottom rows are the no-phase transition, first-, and second-order phase transition injections.  

Figure~\ref{fig:population_probability} tells an interesting story. Looking at the middle panel, we see our method is easily able to distinguish a first-order phase transition as the order-level posterior for the $m=1$ recovery (purple curve) rises quickly and plateaus as a function of the number of events. At the same time, the $m=$none, 2, and 3 posteriors quickly fall to zero (remembering that the sum of the posteriors must equal unity).  However, the injections of $m=$none and $m=2$ equations of state tells a sadder story; we are unable to definitively show evidence of the correct phase transition.

\begin{figure*}
    \centering
    \includegraphics[width=0.98\textwidth]{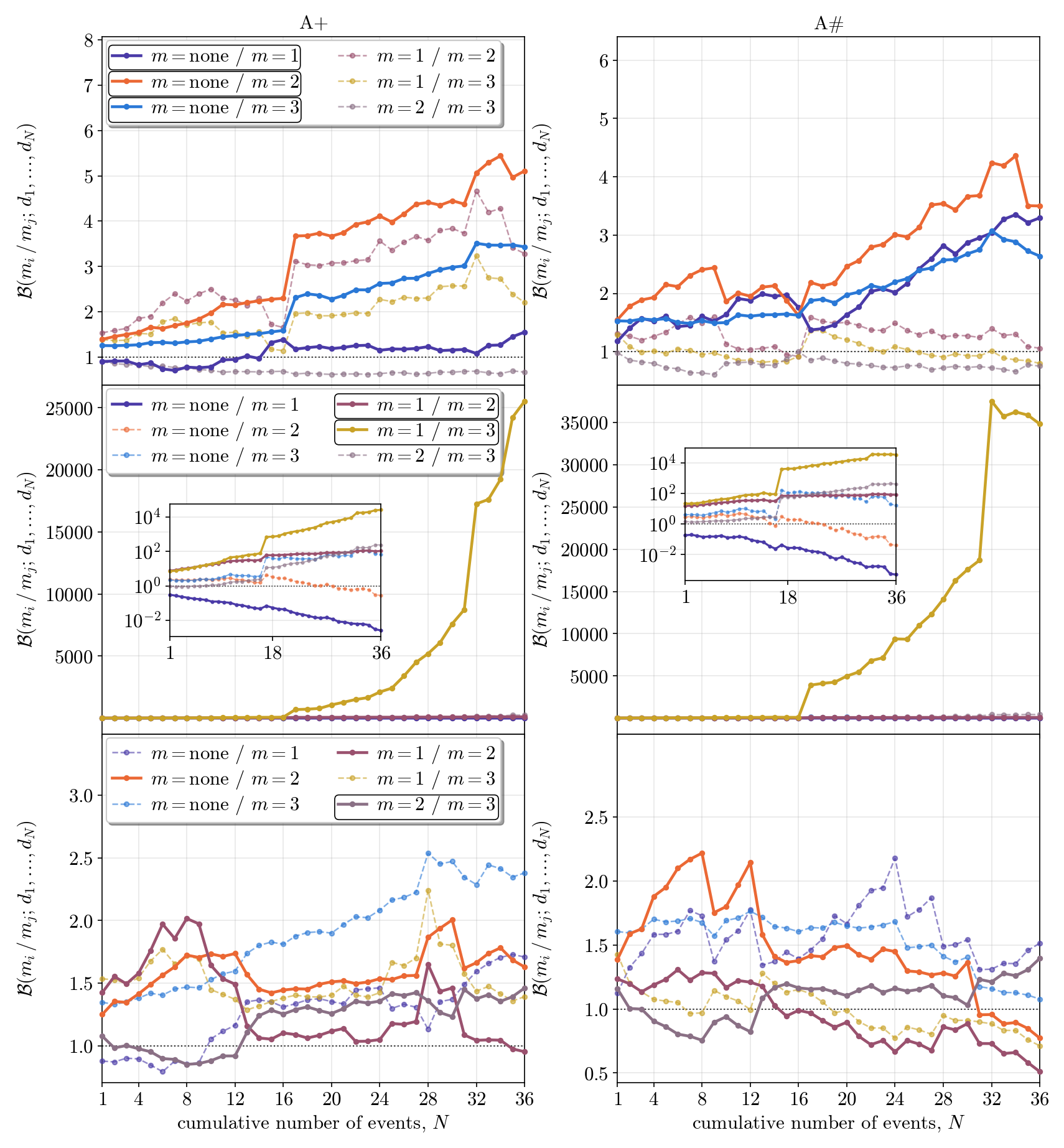}
    \caption{Cumulative Bayes factors between orders. As with Figure~\ref{fig:population_probability}, the left and right columns are injections into A+ and A$^\sharp$ networks, respectively, and the top, middle, and bottom rows show the injections with true equations of state being no phase transition, first-order phase transition and second-order phase transition, respectively.}
    \label{fig:population_BayesFactor}
\end{figure*}

Figure~\ref{fig:population_BayesFactor} shows the corresponding Bayes factors between any given orders, with the same figure ordering as Figure~\ref{fig:population_probability}. The story here is similar: we can easily see the correct modes growing quickly when the true injection is a first-order phase transition ($m=1$; middle row), but this is not the case for the $m=$none and $m=2$ injections.

\begin{figure*}
    \centering
    \includegraphics[width=0.98\textwidth]{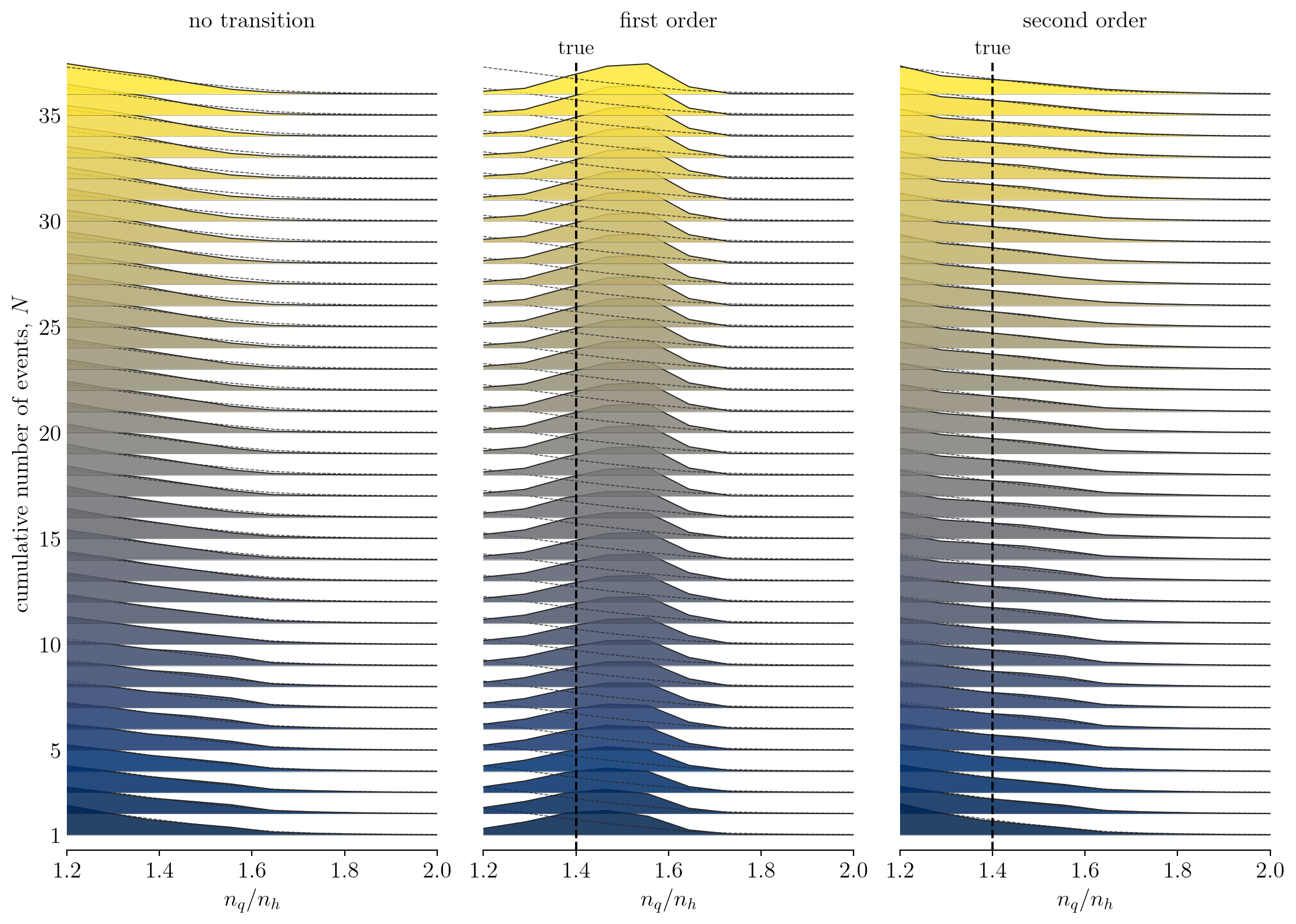}
    \caption{Cumulative marginal distributions $p(n_q / n_h |\left\{d_1\ldots d_N\right\},T)$ for our A$^\sharp$ injections, where the left, middle, and right panel are for universes where the true order of the phase transitions are respectively none, first, and second.}
    \label{fig:Asharp_ridge_nh}
\end{figure*}

Figure~\ref{fig:Asharp_ridge_nh} shows the one-dimensional cumulative marginal distributions for the density jump $n_q / n_h$, where the injections are no phase transition (left panel), a first-order phase transition (middle panel) and a second-order phase transition. To be clear, each value of $N$ shows a cumulative posterior distribution from Equation~\ref{eq:marginal_pop_mixture}, conditioned on the existence of a transition: i.e., $p(n_q / n_h |\left\{d_1\ldots d_N\right\},T)$. We choose to only show the density jump as this is the parameter that was consistently more informative than the others for the single-event analyses, and it is still true in the population analyses that those other parameters are relatively uninformative

Figure~\ref{fig:Asharp_ridge_nh} shows that when there is not a phase transition (left column), the search recovers the priors, however when there is a true first-order phase transition (middle panel), the search recovers that values (albeit with a broad posterior).  We note that in the case of a true second-order phase transition (right panel), the search cannot pick up this parameter, and the marginal posterior is again consistent with the prior.

\begin{figure}
    \centering
    \includegraphics[width=0.98\columnwidth]{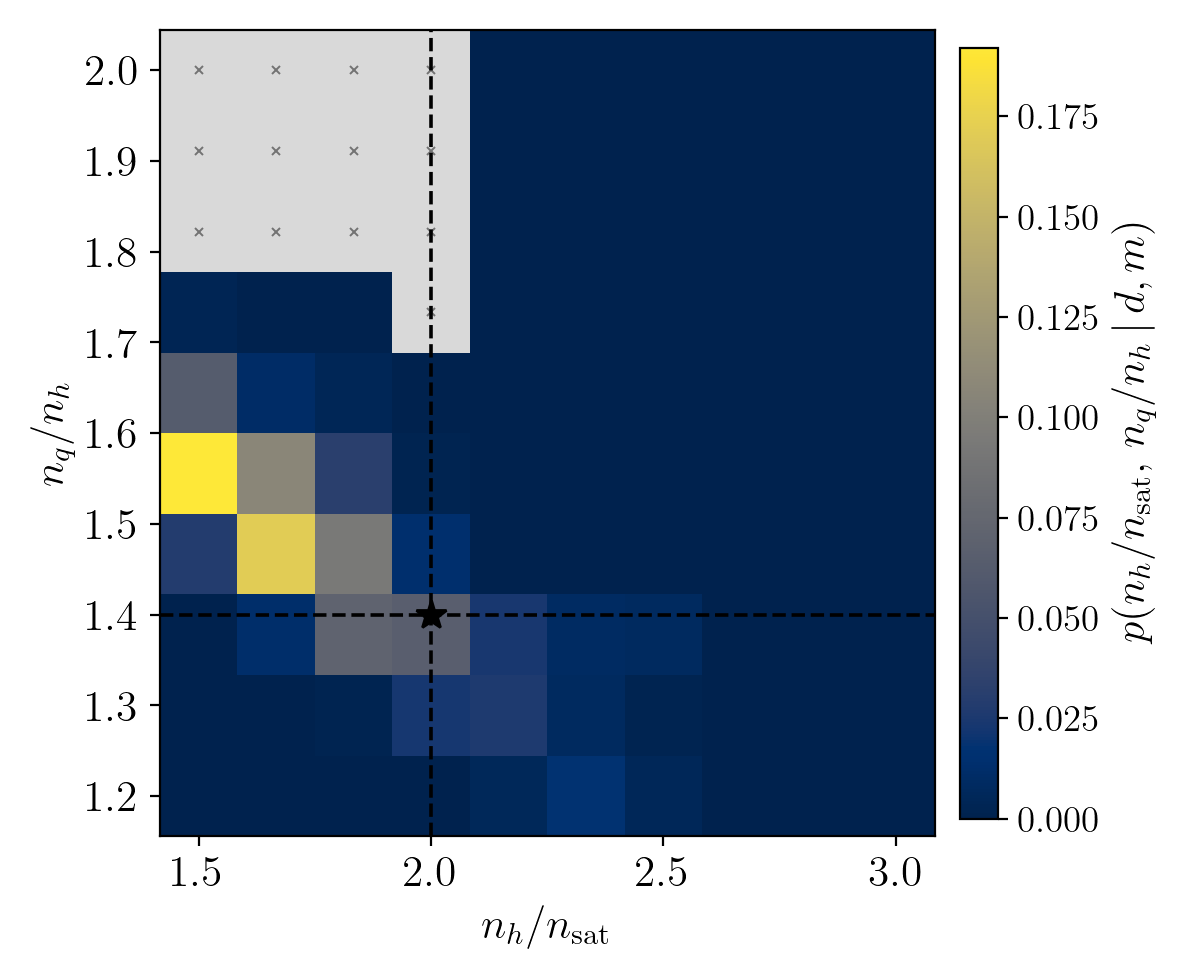}
    \caption{Marginalised posterior for density onset $n_h / n_{\rm sat}$ and density jump $n_q / n_h$ parameters for the A$^\sharp$ synthetic population after 36 mergers.}
    \label{fig:joint_pop_asharp}
\end{figure}

It is worth pointing out that the density jump parameter recovery is covariant with the density onset $n_h$. Figure~\ref{fig:joint_pop_asharp} shows the two-dimensional marginalised posterior for these parameters for 36 events for the $m=1$ Universe, the injection shown as the star, and the colours showing the value of the posterior probablity per cell.  

Despite the `blocky' nature of Figure~\ref{fig:joint_pop_asharp} (due to the coarse nature of our grid as discussed in Section~\ref{sec:phase_transitions} and Table~\ref{tab:transition_grids}), we see the covariance between the two parameters such that small values of $n_h/n_{\rm sat}$ correspond to larger values of $n_q/n_h$, which our model is unable to distinguish. This is \textit{not} entirely due to the fact that both axes contain the $n_h$ parameter; when we plot a posterior of $n_h / n_{\rm sat}$ against $n_q / n_{\rm sat}$, for example, we still see a covariance, albeit with a different shape. 
Whichever parameter space we wish to work in, marginalising out $n_q/n_h$ leads to an apparent bias in $n_{h}/n_{\rm sat}$ (not shown).

The summary of this section is therefore somewhat mixed! If a universe has a true first-order phase transition (with relatively low transition density), then we will be able to definitively measure it with multiple neutron star mergers, and also get a handle on the density jump ratio parameter, but no other parameters. If, on the other hand, there is no phase transition, or there is a second-order phase transition, or there is a first-order phase transition with relatively high transition density, we will not be able to distinguish these three scenarios from the inspiral. In any case, it will be difficult to measure the transition onset density.

\section{Conclusions}

Despite a somewhat lacklustre period for binary neutron star gravitational-wave observations, the future of the field is positive if the LIGO-Virgo-KAGRA observatories can reach their ultimate design sensitivities and beyond. With those observations comes the opportunity to probe nuclear matter at supranuclear densities, including the possibility of measuring QCD phase transitions in neutron star cores. In this work, we extend this to the \textit{order} of possible phase transitions.

Utilising recent work in extending hadronic equations of state to include first-, second-, and third-order phase transitions~\citep[e.g.,][]{alford13,clevinger26}, we construct hybrid equations of state of all three orders that include hadronic low densities and deconfined quark states at high densities. We develop the statistical framework to test how consistent these large families of equations of state are with gravitational-wave data from binary neutron star inspiral signals. We also develop the population-level statistics to determine the consistency of equations of state from ensembles of such gravitational-wave observations.
      
We show that the binary neutron star mergers GW170817 and GW190425 are relatively uninformative, although GW170817 shows some evidence for a first-order phase transition consistent, qualitatively consistent with earlier works~\citep{essick20,blaschke20,annala20,cao26}. If anything, it also has a small preference for a first-order phase transition. GW190425 is, unsurprisingly, consistent with everything all at once. We simulate binary neutron star mergers and inject them into data for observatories at A+, A$^\sharp$ and Cosmic Explorer sensitivity. The Cosmic Explorer injection, in particular, shows that a phase transition order can be measured; we simulated a first-order phase transition and showed the order-level model was recovered with posterior probability of 74\%, compared to all other orders being preferred at most 10\% each. Of course, this was for a single injection with specific parameters, so one can expect this to change depending on the merger itself as well as the possible nature of the phase transition.

Our method is considerably more powerful at the population level when trying to measure phase transition orders from ensembles of mergers. We performed synthetic injections into networks of both A+ and A$^\sharp$ observatories, showing that first-order phase transitions are recovered with $p(m=1)>0.9$ after about 14 events in both A+ and A$^\sharp$. This is consistent, albeit at the optimistic end, with other works~\citep{pang20,chatziioannou20,essick23}. 
Unfortunately, though, second- and third-order phase transitions are significantly more difficult, and they can also easily be mistaken for one another or when there is no phase transition present at all.
It is worth stressing our results are also dependent on our specific choices of baseline equations of state and phase-transition parameters, all of which, in turn, define our priors. We leave it as future work to explore the robustness of those modelling assumptions.

There are a number of uncertainties and modelling assumptions that we have made along the way that could have a significant impact on our conclusions. For example, for our first-order phase transitions in our synthetic population, we chose parameters such that every neutron stars more massive than $0.9\,M_\odot$ would contain quark matter in its core. In other words, \textit{all} stars in our population are hybrid stars! If, on the other hand, the onset density were much larger, this would harm our ability to measure such a phase transition. 

Similarly to the above, it is worth again emphasising that the mass distribution of neutron stars is unknown, in large part due to the unusually massive (when compared to the population of observed Galactic double neutron stars) system GW190425. We attempt to take this into account by modelling the mass distribution as a double-peaked Gaussian, but the parameters of that heavier peak are highly uncertain. On the other hand, it is exactly the heavier neutron stars that are better able to probe phase transitions due to their higher central densities. If the Universe is unkind, and GW190425 is not representative of merging neutron stars, then it may be more difficult to measure phase transitions as outlined in this work.

While several aspects of this study may make our results optimistic, several avenues could also improve the prospects of measuring phase transitions over what is described herein. Three things that come to mind:
\begin{itemize}
    \item We have not robustly considered other astrophysical constraints, such as neutron star radius measurements from the NICER telescope~\citep{riley19,miller19,riley21,miller21,choudhury24}. Combining these with gravitational-wave observations~\citep[e.g.,][]{raaijmakers21,legred21} will provide more stringent constraints, a task we leave for future work.
    \item We have not included potential electromagnetic counterparts, which measure e.g., the ejecta mass and kilonova colour, and hence provide further equation-of-state information~\citep[e.g.,][]{margalit17,radice18,coughlin19,dietrich20}.
    \item We have only included the gravitational-wave inspiral portion of the signal; however the high-frequency post-merger signal can also be observed, which would give complementary information about the high-temperature equation of state~\citep[e.g.,][]{bauswein12,takami14,bernuzzi15}, as well as provide the ability to measure temperature-dependent phase transitions~\citep[e.g.,][]{most19,bauswein19,weih20,fujimoto23}.
\end{itemize}

\section*{Acknowledgements}
PL used Claude (Anthropic) and Gemini (Google) for code generation, script optimisation and debugging, as well as text massaging, which includes the drafting of this sentence. We are grateful to Nir Guttman for help with the Cosmic Explorer simulation, and Debarati Chatterjee for helpful comments on the manuscript.

PL is supported by Australian Research Council grants ARC LE210100002, DP230103088, and CE230100016. This work was performed on the OzSTAR national facility at Swinburne University of Technology. The OzSTAR program receives funding in part from the Astronomy National Collaborative Research Infrastructure Strategy (NCRIS) allocation provided by the Australian Government, and from the Victorian Higher Education State Investment Fund (VHESIF) provided by the Victorian Government. This work was partially supported by Conselho Nacional de Desenvolvimento Cient\'ifico e Tecno\-l\'o\-gico  (CNPq), Grants No. 312032/2023-4, No. 402963/2024-5 and 445182/2024-5 (R.L.S.F.),  200037/2026-9 (W.R.T.) and 315225/2025-4 (T.F.M.); Funda\c{c}\~ao de Amparo \`a Pesquisa do Estado do Rio Grande do Sul (FAPERGS), Grants No. 24/2551-0001285-0 (R.L.S.F.). CAPES Finance Code 001 (A.G.S). The work is also part of the project Instituto Nacional de Ci\^encia e Tecnologia - F\'isica Nuclear e Aplica\c{c}\~oes (INCT - FNA), Grants No. 464898/2014-5 and No. 408419/2024-5, and supported by the Ser\-ra\-pi\-lhei\-ra Institute (Grant No. Serra - 2211-42230). AGS acknowledges support from Coordena\c{c}\~ao de Aperfei\c{c}oamento de Pessoal de N\'ivel Superior (CAPES), Finance Code 001. RLSF acknowledges the kind hospitality of the Center for Nuclear Research at Kent State University, where part of this work was done. VD acknowledges support from the U.S. Department of Energy under grant DE-SC0024700, from the National Science Foundation under grants MUSES OAC2103680 and NP3M PHY-2116686, PHY-2621752, PHY-2623480, and from the Simons Foundation under award Pivot Fellow-00016159.

\bibliographystyle{mnras}
\bibliography{Bib}

\begin{thebibliography}{}
\makeatletter
\relax
\def\mn@urlcharsother{\let\do\@makeother \do\$\do\&\do\#\do\^\do\_\do\%\do\~}
\def\mn@doi{\begingroup\mn@urlcharsother \@ifnextchar [ {\mn@doi@}
  {\mn@doi@[]}}
\def\mn@doi@[#1]#2{\def\@tempa{#1}\ifx\@tempa\@empty \href
  {http://dx.doi.org/#2} {doi:#2}\else \href {http://dx.doi.org/#2} {#1}\fi
  \endgroup}
\def\mn@eprint#1#2{\mn@eprint@#1:#2::\@nil}
\def\mn@eprint@arXiv#1{\href {http://arxiv.org/abs/#1} {{\tt arXiv:#1}}}
\def\mn@eprint@dblp#1{\href {http://dblp.uni-trier.de/rec/bibtex/#1.xml}
  {dblp:#1}}
\def\mn@eprint@#1:#2:#3:#4\@nil{\def\@tempa {#1}\def\@tempb {#2}\def\@tempc
  {#3}\ifx \@tempc \@empty \let \@tempc \@tempb \let \@tempb \@tempa \fi \ifx
  \@tempb \@empty \def\@tempb {arXiv}\fi \@ifundefined
  {mn@eprint@\@tempb}{\@tempb:\@tempc}{\expandafter \expandafter \csname
  mn@eprint@\@tempb\endcsname \expandafter{\@tempc}}}

\bibitem[\protect\citeauthoryear{{Aasi}, {Abbott}, {Abbott}, {Abbott},
  {Abernathy}  et~al.}{{Aasi} et~al.}{2015}]{aasi15}
{Aasi} J.,  {Abbott} B.~P.,  {Abbott} R.,  {Abbott} T.,  {Abernathy} M.~R.,
  et~al., 2015, \mn@doi [Classical and Quantum Gravity]
  {10.1088/0264-9381/32/7/074001}, \href
  {https://ui.adsabs.harvard.edu/abs/2015CQGra..32g4001L} {32, 074001}

\bibitem[\protect\citeauthoryear{{Abbott} et~al.}{{Abbott}
  et~al.}{2017a}]{abbott17_CE}
{Abbott} B.~P.,  et~al., 2017a, \mn@doi [Classical and Quantum Gravity]
  {10.1088/1361-6382/aa51f4}, \href
  {https://ui.adsabs.harvard.edu/abs/2017CQGra..34d4001A} {34, 044001}

\bibitem[\protect\citeauthoryear{{Abbott} et~al.}{{Abbott}
  et~al.}{2017b}]{abbott17_GW170817detection}
{Abbott} B.~P.,  et~al., 2017b, \mn@doi [\prl]
  {10.1103/PhysRevLett.119.161101}, \href
  {https://ui.adsabs.harvard.edu/abs/2017PhRvL.119p1101A} {119, 161101}

\bibitem[\protect\citeauthoryear{{Abbott} et~al.}{{Abbott}
  et~al.}{2017c}]{abbott17_GW170817MMA}
{Abbott} B.~P.,  et~al., 2017c, \mn@doi [\apjl] {10.3847/2041-8213/aa91c9},
  \href {https://ui.adsabs.harvard.edu/abs/2017ApJ...848L..12A} {848, L12}

\bibitem[\protect\citeauthoryear{{Abbott}, {Abbott}, {Abbott}  et~al.}{{Abbott}
  et~al.}{2018a}]{abbott18_Aplus}
{Abbott} B.~P.,  {Abbott} R.,  {Abbott} T.~D.,   et~al., 2018a, \mn@doi [Living
  Reviews in Relativity] {10.1007/s41114-018-0012-9}, \href
  {https://ui.adsabs.harvard.edu/abs/2018LRR....21....3A} {21, 3}

\bibitem[\protect\citeauthoryear{{Abbott} et~al.}{{Abbott}
  et~al.}{2018b}]{abbott18_GW170817nsradii}
{Abbott} B.~P.,  et~al., 2018b, \mn@doi [\prl]
  {10.1103/PhysRevLett.121.161101}, \href
  {https://ui.adsabs.harvard.edu/abs/2018PhRvL.121p1101A} {121, 161101}

\bibitem[\protect\citeauthoryear{{Abbott} et~al.}{{Abbott}
  et~al.}{2019}]{abbott19_gwtc1}
{Abbott} B.~P.,  et~al., 2019, \mn@doi [Physical Review X]
  {10.1103/PhysRevX.9.031040}, \href
  {https://ui.adsabs.harvard.edu/abs/2019PhRvX...9c1040A} {9, 031040}

\bibitem[\protect\citeauthoryear{{Abbott} et~al.}{{Abbott}
  et~al.}{2020}]{abbott20_GW190425}
{Abbott} B.~P.,  et~al., 2020, \mn@doi [\apjl] {10.3847/2041-8213/ab75f5},
  \href {https://ui.adsabs.harvard.edu/abs/2020ApJ...892L...3A} {892, L3}

\bibitem[\protect\citeauthoryear{{Acernese} et~al.}{{Acernese}
  et~al.}{2015}]{acernese15}
{Acernese} F.,  et~al., 2015, \mn@doi [Classical and Quantum Gravity]
  {10.1088/0264-9381/32/2/024001}, \href
  {https://ui.adsabs.harvard.edu/abs/2015CQGra..32b4001A} {32, 024001}

\bibitem[\protect\citeauthoryear{{Agrawal}, {Shlomo}  \& {Au}}{{Agrawal}
  et~al.}{2003}]{agrawal03}
{Agrawal} B.~K.,  {Shlomo} S.,   {Au} V.~K.,  2003, \mn@doi [Phys. Rev. C]
  {10.1103/PhysRevC.68.031304}, \href
  {https://ui.adsabs.harvard.edu/abs/2003PhRvC..68c1304A} {68, 031304}

\bibitem[\protect\citeauthoryear{Akutsu et~al.}{Akutsu et~al.}{2021}]{akutsu21}
Akutsu T.,  et~al., 2021, \mn@doi [Progress of Theoretical and Experimental
  Physics] {10.1093/ptep/ptaa125}, 2021, 05A101

\bibitem[\protect\citeauthoryear{{Alford}, {Han}  \& {Prakash}}{{Alford}
  et~al.}{2013}]{alford13}
{Alford} M.~G.,  {Han} S.,   {Prakash} M.,  2013, \mn@doi [Phys. Rev. D]
  {10.1103/PhysRevD.88.083013}, \href
  {https://ui.adsabs.harvard.edu/abs/2013PhRvD..88h3013A} {88, 083013}

\bibitem[\protect\citeauthoryear{{Annala}, {Gorda}, {Kurkela}  \&
  {Vuorinen}}{{Annala} et~al.}{2018}]{annala18}
{Annala} E.,  {Gorda} T.,  {Kurkela} A.,   {Vuorinen} A.,  2018, \mn@doi [\prl]
  {10.1103/PhysRevLett.120.172703}, \href
  {https://ui.adsabs.harvard.edu/abs/2018PhRvL.120q2703A} {120, 172703}

\bibitem[\protect\citeauthoryear{{Annala}, {Gorda}, {Kurkela},
  {N{\"a}ttil{\"a}}  \& {Vuorinen}}{{Annala} et~al.}{2020}]{annala20}
{Annala} E.,  {Gorda} T.,  {Kurkela} A.,  {N{\"a}ttil{\"a}} J.,   {Vuorinen}
  A.,  2020, \mn@doi [Nature Physics] {10.1038/s41567-020-0914-9}, \href
  {https://ui.adsabs.harvard.edu/abs/2020NatPh..16..907A} {16, 907}

\bibitem[\protect\citeauthoryear{{Ashton} et~al.,}{{Ashton}
  et~al.}{2019}]{ashton19}
{Ashton} G.,  et~al., 2019, \mn@doi [\apjs] {10.3847/1538-4365/ab06fc}, \href
  {https://ui.adsabs.harvard.edu/abs/2019ApJS..241...27A} {241, 27}

\bibitem[\protect\citeauthoryear{{Baker}, {Lasky}, {Thrane}  \&
  {Golomb}}{{Baker} et~al.}{2025}]{baker25}
{Baker} A.~M.,  {Lasky} P.~D.,  {Thrane} E.,   {Golomb} J.,  2025, \mn@doi
  [\prd] {10.1103/n6t6-5wn3}, \href
  {https://ui.adsabs.harvard.edu/abs/2025PhRvD.112j2004B} {112, 102004}

\bibitem[\protect\citeauthoryear{{Bauswein} \& {Janka}}{{Bauswein} \&
  {Janka}}{2012}]{bauswein12}
{Bauswein} A.,  {Janka} H.-T.,  2012, \mn@doi [\prl]
  {10.1103/PhysRevLett.108.011101}, \href
  {https://ui.adsabs.harvard.edu/abs/2012PhRvL.108a1101B} {108, 011101}

\bibitem[\protect\citeauthoryear{{Bauswein}, {Bastian}, {Blaschke},
  {Chatziioannou}, {Clark}, {Fischer}  \& {Oertel}}{{Bauswein}
  et~al.}{2019}]{bauswein19}
{Bauswein} A.,  {Bastian} N.-U.~F.,  {Blaschke} D.~B.,  {Chatziioannou} K.,
  {Clark} J.~A.,  {Fischer} T.,   {Oertel} M.,  2019, \mn@doi [\prl]
  {10.1103/PhysRevLett.122.061102}, \href
  {https://ui.adsabs.harvard.edu/abs/2019PhRvL.122f1102B} {122, 061102}

\bibitem[\protect\citeauthoryear{{Baym}, {Hatsuda}, {Kojo}, {Powell}, {Song}
  \& {Takatsuka}}{{Baym} et~al.}{2018}]{baym18}
{Baym} G.,  {Hatsuda} T.,  {Kojo} T.,  {Powell} P.~D.,  {Song} Y.,
  {Takatsuka} T.,  2018, \mn@doi [Reports on Progress in Physics]
  {10.1088/1361-6633/aaae14}, \href
  {https://ui.adsabs.harvard.edu/abs/2018RPPh...81e6902B} {81, 056902}

\bibitem[\protect\citeauthoryear{{Bednarek}, {Haensel}, {Zdunik}, {Bejger}  \&
  {Ma{\'n}ka}}{{Bednarek} et~al.}{2012}]{bednarek12}
{Bednarek} I.,  {Haensel} P.,  {Zdunik} J.~L.,  {Bejger} M.,   {Ma{\'n}ka} R.,
  2012, \mn@doi [\aap] {10.1051/0004-6361/201118560}, \href
  {https://ui.adsabs.harvard.edu/abs/2012A&A...543A.157B} {543, A157}

\bibitem[\protect\citeauthoryear{{Bernuzzi}, {Dietrich}  \& {Nagar}}{{Bernuzzi}
  et~al.}{2015}]{bernuzzi15}
{Bernuzzi} S.,  {Dietrich} T.,   {Nagar} A.,  2015, \mn@doi [\prl]
  {10.1103/PhysRevLett.115.091101}, \href
  {https://ui.adsabs.harvard.edu/abs/2015PhRvL.115i1101B} {115, 091101}

\bibitem[\protect\citeauthoryear{{Blaschke}, {Ayriyan}, {Alvarez-Castillo}  \&
  {Grigorian}}{{Blaschke} et~al.}{2020}]{blaschke20}
{Blaschke} D.,  {Ayriyan} A.,  {Alvarez-Castillo} D.~E.,   {Grigorian} H.,
  2020, \mn@doi [Universe] {10.3390/universe6060081}, \href
  {https://ui.adsabs.harvard.edu/abs/2020Univ....6...81B} {6, 81}

\bibitem[\protect\citeauthoryear{{Branchesi} et~al.}{{Branchesi}
  et~al.}{2023}]{branchesi23}
{Branchesi} M.,  et~al., 2023, \mn@doi [\jcap] {10.1088/1475-7516/2023/07/068},
  \href {https://ui.adsabs.harvard.edu/abs/2023JCAP...07..068B} {2023, 068}

\bibitem[\protect\citeauthoryear{{Burgio}, {Schulze}, {Vida{\~n}a}  \&
  {Wei}}{{Burgio} et~al.}{2021}]{burgio21}
{Burgio} G.~F.,  {Schulze} H.-J.,  {Vida{\~n}a} I.,   {Wei} J.-B.,  2021,
  \mn@doi [Progress in Particle and Nuclear Physics]
  {10.1016/j.ppnp.2021.103879}, \href
  {https://ui.adsabs.harvard.edu/abs/2021PrPNP.12003879B} {120, 103879}

\bibitem[\protect\citeauthoryear{{Cao} \& {Chen}}{{Cao} \&
  {Chen}}{2026}]{cao26}
{Cao} Z.,  {Chen} L.-W.,  2026, \mn@doi [\apj] {10.3847/1538-4357/ae946d},
  \href {https://ui.adsabs.harvard.edu/abs/2026ApJ..1008..211C} {1008, 211}

\bibitem[\protect\citeauthoryear{{Chatterjee} \& {Vida{\~n}a}}{{Chatterjee} \&
  {Vida{\~n}a}}{2016}]{chatterjee16}
{Chatterjee} D.,  {Vida{\~n}a} I.,  2016, \mn@doi [European Physical Journal A]
  {10.1140/epja/i2016-16029-x}, \href
  {https://ui.adsabs.harvard.edu/abs/2016EPJA...52...29C} {52, 29}

\bibitem[\protect\citeauthoryear{{Chatziioannou} \& {Han}}{{Chatziioannou} \&
  {Han}}{2020}]{chatziioannou20}
{Chatziioannou} K.,  {Han} S.,  2020, \mn@doi [\prd]
  {10.1103/PhysRevD.101.044019}, \href
  {https://ui.adsabs.harvard.edu/abs/2020PhRvD.101d4019C} {101, 044019}

\bibitem[\protect\citeauthoryear{{Choudhury} et~al.,}{{Choudhury}
  et~al.}{2024}]{choudhury24}
{Choudhury} D.,  et~al., 2024, \mn@doi [\apjl] {10.3847/2041-8213/ad5a6f},
  \href {https://ui.adsabs.harvard.edu/abs/2024ApJ...971L..20C} {971, L20}

\bibitem[\protect\citeauthoryear{{Clevinger}, {Corkish}, {Aryal}  \&
  {Dexheimer}}{{Clevinger} et~al.}{2022}]{clevinger22}
{Clevinger} A.,  {Corkish} J.,  {Aryal} K.,   {Dexheimer} V.,  2022, \mn@doi
  [European Physical Journal A] {10.1140/epja/s10050-022-00745-3}, \href
  {https://ui.adsabs.harvard.edu/abs/2022EPJA...58...96C} {58, 96}

\bibitem[\protect\citeauthoryear{{Clevinger}, {Lin}, {Albino}, {Hammond},
  {Steiner}  \& {Dexheimer}}{{Clevinger} et~al.}{2026}]{clevinger26}
{Clevinger} A.,  {Lin} Z.,  {Albino} M.,  {Hammond} P.,  {Steiner} A.,
  {Dexheimer} V.,  2026, \mn@doi [\prd] {10.1103/32rl-1lwc}, \href
  {https://ui.adsabs.harvard.edu/abs/2026PhRvD.113h1301C} {113, L081301}

\bibitem[\protect\citeauthoryear{{Collins} \& {Perry}}{{Collins} \&
  {Perry}}{1975}]{collins75}
{Collins} J.~C.,  {Perry} M.~J.,  1975, \mn@doi [\prl]
  {10.1103/PhysRevLett.34.1353}, \href
  {https://ui.adsabs.harvard.edu/abs/1975PhRvL..34.1353C} {34, 1353}

\bibitem[\protect\citeauthoryear{{Coughlin}, {Dietrich}, {Margalit}  \&
  {Metzger}}{{Coughlin} et~al.}{2019}]{coughlin19}
{Coughlin} M.~W.,  {Dietrich} T.,  {Margalit} B.,   {Metzger} B.~D.,  2019,
  \mn@doi [\mnras] {10.1093/mnrasl/slz133}, \href
  {https://ui.adsabs.harvard.edu/abs/2019MNRAS.489L..91C} {489, L91}

\bibitem[\protect\citeauthoryear{{Cromartie} et~al.,}{{Cromartie}
  et~al.}{2020}]{cromartie20}
{Cromartie} H.~T.,  et~al., 2020, \mn@doi [Nature Astronomy]
  {10.1038/s41550-019-0880-2}, \href
  {https://ui.adsabs.harvard.edu/abs/2020NatAs...4...72C} {4, 72}

\bibitem[\protect\citeauthoryear{{Damour} \& {Nagar}}{{Damour} \&
  {Nagar}}{2009}]{damour09}
{Damour} T.,  {Nagar} A.,  2009, \mn@doi [\prd] {10.1103/PhysRevD.80.084035},
  \href {https://ui.adsabs.harvard.edu/abs/2009PhRvD..80h4035D} {80, 084035}

\bibitem[\protect\citeauthoryear{{Danielewicz} \& {Lee}}{{Danielewicz} \&
  {Lee}}{2009}]{danielewicz09}
{Danielewicz} P.,  {Lee} J.,  2009, \mn@doi [\nphysa]
  {10.1016/j.nuclphysa.2008.11.007}, \href
  {https://ui.adsabs.harvard.edu/abs/2009NuPhA.818...36D} {818, 36}

\bibitem[\protect\citeauthoryear{{Dietrich}, {Bernuzzi}  \& {Tichy}}{{Dietrich}
  et~al.}{2017}]{dietrich17}
{Dietrich} T.,  {Bernuzzi} S.,   {Tichy} W.,  2017, \mn@doi [\prd]
  {10.1103/PhysRevD.96.121501}, \href
  {https://ui.adsabs.harvard.edu/abs/2017PhRvD..96l1501D} {96, 121501}

\bibitem[\protect\citeauthoryear{{Dietrich}, {Coughlin}, {Pang}, {Bulla},
  {Heinzel}, {Issa}, {Tews}  \& {Antier}}{{Dietrich} et~al.}{2020}]{dietrich20}
{Dietrich} T.,  {Coughlin} M.~W.,  {Pang} P. T.~H.,  {Bulla} M.,  {Heinzel} J.,
   {Issa} L.,  {Tews} I.,   {Antier} S.,  2020, \mn@doi [Science]
  {10.1126/science.abb4317}, \href
  {https://ui.adsabs.harvard.edu/abs/2020Sci...370.1450D} {370, 1450}

\bibitem[\protect\citeauthoryear{{Essick}, {Landry}  \& {Holz}}{{Essick}
  et~al.}{2020}]{essick20}
{Essick} R.,  {Landry} P.,   {Holz} D.~E.,  2020, \mn@doi [\prd]
  {10.1103/PhysRevD.101.063007}, \href
  {https://ui.adsabs.harvard.edu/abs/2020PhRvD.101f3007E} {101, 063007}

\bibitem[\protect\citeauthoryear{{Essick}, {Legred}, {Chatziioannou}, {Han}  \&
  {Landry}}{{Essick} et~al.}{2023}]{essick23}
{Essick} R.,  {Legred} I.,  {Chatziioannou} K.,  {Han} S.,   {Landry} P.,
  2023, \mn@doi [\prd] {10.1103/PhysRevD.108.043013}, \href
  {https://ui.adsabs.harvard.edu/abs/2023PhRvD.108d3013E} {108, 043013}

\bibitem[\protect\citeauthoryear{{Evans} et~al.}{{Evans}
  et~al.}{2021}]{evans21}
{Evans} M.,  et~al., 2021, \mn@doi [arXiv e-prints]
  {10.48550/arXiv.2109.09882}, \href
  {https://ui.adsabs.harvard.edu/abs/2021arXiv210909882E} {p. arXiv:2109.09882}

\bibitem[\protect\citeauthoryear{{Farrow}, {Zhu}  \& {Thrane}}{{Farrow}
  et~al.}{2019}]{farrow19}
{Farrow} N.,  {Zhu} X.-J.,   {Thrane} E.,  2019, \mn@doi [\apj]
  {10.3847/1538-4357/ab12e3}, \href
  {https://ui.adsabs.harvard.edu/abs/2019ApJ...876...18F} {876, 18}

\bibitem[\protect\citeauthoryear{{Fattoyev}, {Piekarewicz}  \&
  {Horowitz}}{{Fattoyev} et~al.}{2018}]{fattoyev18}
{Fattoyev} F.~J.,  {Piekarewicz} J.,   {Horowitz} C.~J.,  2018, \mn@doi [\prl]
  {10.1103/PhysRevLett.120.172702}, \href
  {https://ui.adsabs.harvard.edu/abs/2018PhRvL.120q2702F} {120, 172702}

\bibitem[\protect\citeauthoryear{{Flanagan} \& {Hinderer}}{{Flanagan} \&
  {Hinderer}}{2008}]{flanagan08}
{Flanagan} {\'E}.~{\'E}.,  {Hinderer} T.,  2008, \mn@doi [\prd]
  {10.1103/PhysRevD.77.021502}, \href
  {https://ui.adsabs.harvard.edu/abs/2008PhRvD..77b1502F} {77, 021502}

\bibitem[\protect\citeauthoryear{{Fonseca} et~al.,}{{Fonseca}
  et~al.}{2021}]{fonseca21}
{Fonseca} E.,  et~al., 2021, \mn@doi [Astrophysical Journal Letters]
  {10.3847/2041-8213/ac03b8}, \href
  {https://ui.adsabs.harvard.edu/abs/2021ApJ...915L..12F} {915, L12}

\bibitem[\protect\citeauthoryear{{Fraga} \& {Romatschke}}{{Fraga} \&
  {Romatschke}}{2005}]{fraga05}
{Fraga} E.~S.,  {Romatschke} P.,  2005, \mn@doi [\prd]
  {10.1103/PhysRevD.71.105014}, \href
  {https://ui.adsabs.harvard.edu/abs/2005PhRvD..71j5014F} {71, 105014}

\bibitem[\protect\citeauthoryear{{Freedman} \& {McLerran}}{{Freedman} \&
  {McLerran}}{1977}]{freedman77}
{Freedman} B.~A.,  {McLerran} L.~D.,  1977, \mn@doi [\prd]
  {10.1103/PhysRevD.16.1169}, \href
  {https://ui.adsabs.harvard.edu/abs/1977PhRvD..16.1169F} {16, 1169}

\bibitem[\protect\citeauthoryear{{Fritschel}, {Kabagoz}, {Kuns}  \&
  {Effler}}{{Fritschel} et~al.}{2025}]{fritschel25}
{Fritschel} P.,  {Kabagoz} B.,  {Kuns} K.,   {Effler} A.,  2025, Technical
  Report LIGO-T2500310, {A+/O5 strain curve projections}, \url
  {https://dcc.ligo.org/LIGO-T2500310/public}.
LIGO, \url {https://dcc.ligo.org/LIGO-T2500310/public}

\bibitem[\protect\citeauthoryear{{Fujimoto}, {Fukushima}, {Hotokezaka}  \&
  {Kyutoku}}{{Fujimoto} et~al.}{2023}]{fujimoto23}
{Fujimoto} Y.,  {Fukushima} K.,  {Hotokezaka} K.,   {Kyutoku} K.,  2023,
  \mn@doi [\prl] {10.1103/PhysRevLett.130.091404}, \href
  {https://ui.adsabs.harvard.edu/abs/2023PhRvL.130i1404F} {130, 091404}

\bibitem[\protect\citeauthoryear{{Ghiglieri}, {Kurkela}, {Strickland}  \&
  {Vuorinen}}{{Ghiglieri} et~al.}{2020}]{ghiglieri20}
{Ghiglieri} J.,  {Kurkela} A.,  {Strickland} M.,   {Vuorinen} A.,  2020,
  \mn@doi [\physrep] {10.1016/j.physrep.2020.07.004}, \href
  {https://ui.adsabs.harvard.edu/abs/2020PhR...880....1G} {880, 1}

\bibitem[\protect\citeauthoryear{{Gulminelli} \& {Raduta}}{{Gulminelli} \&
  {Raduta}}{2015}]{gulminelli15}
{Gulminelli} F.,  {Raduta} A.~R.,  2015, \mn@doi [\prc]
  {10.1103/PhysRevC.92.055803}, \href
  {https://ui.adsabs.harvard.edu/abs/2015PhRvC..92e5803G} {92, 055803}

\bibitem[\protect\citeauthoryear{Guttman, Baker, Lasky  \& Thrane}{Guttman
  et~al.}{2026}]{guttman26}
Guttman N.,  Baker A.~M.,  Lasky P.~D.,   Thrane E.,  2026, \mn@doi [Phys. Rev.
  D] {10.1103/95x8-csn6}, 114, 063035

\bibitem[\protect\citeauthoryear{{Hernandez Vivanco}, {Smith}, {Thrane},
  {Lasky}, {Talbot}  \& {Raymond}}{{Hernandez Vivanco}
  et~al.}{2019}]{hernandezvivanco19}
{Hernandez Vivanco} F.,  {Smith} R.,  {Thrane} E.,  {Lasky} P.~D.,  {Talbot}
  C.,   {Raymond} V.,  2019, \mn@doi [\prd] {10.1103/PhysRevD.100.103009},
  \href {https://ui.adsabs.harvard.edu/abs/2019PhRvD.100j3009H} {100, 103009}

\bibitem[\protect\citeauthoryear{{Hinderer}}{{Hinderer}}{2008}]{hinderer08}
{Hinderer} T.,  2008, \mn@doi [\apj] {10.1086/533487}, \href
  {https://ui.adsabs.harvard.edu/abs/2008ApJ...677.1216H} {677, 1216}

\bibitem[\protect\citeauthoryear{{Iyer}, {Souradeep}, {Unnikrishnan},
  {Dhurandhar}, {Raja}  \& {Sengupta}}{{Iyer} et~al.}{2011}]{iyer11}
{Iyer} B.,  {Souradeep} T.,  {Unnikrishnan} C.~S.,  {Dhurandhar} S.,  {Raja}
  S.,   {Sengupta} A.,  2011, Technical Report LIGO-M1100296, {LIGO-India,
  Proposal of the Consortium for Indian Initiative in Gravitational-wave
  Observations (IndIGO)}, \url {https://dcc.ligo.org/LIGO-M1100296/public}.
LIGO, \url {https://dcc.ligo.org/LIGO-M1100296/public}

\bibitem[\protect\citeauthoryear{{Kojo}, {Powell}, {Song}  \& {Baym}}{{Kojo}
  et~al.}{2015}]{kojo15}
{Kojo} T.,  {Powell} P.~D.,  {Song} Y.,   {Baym} G.,  2015, \mn@doi [\prd]
  {10.1103/PhysRevD.91.045003}, \href
  {https://ui.adsabs.harvard.edu/abs/2015PhRvD..91d5003K} {91, 045003}

\bibitem[\protect\citeauthoryear{{Kumar}, {Dexheimer}  et~al.}{{Kumar}
  et~al.}{2024}]{kumar24}
{Kumar} R.,  {Dexheimer} V.,   et~al., 2024, \mn@doi [Living Reviews in
  Relativity] {10.1007/s41114-024-00049-6}, \href
  {https://ui.adsabs.harvard.edu/abs/2024LRR....27....3K} {27, 3}

\bibitem[\protect\citeauthoryear{{Lackey} \& {Wade}}{{Lackey} \&
  {Wade}}{2015}]{lackey15}
{Lackey} B.~D.,  {Wade} L.,  2015, \mn@doi [\prd] {10.1103/PhysRevD.91.043002},
  \href {https://ui.adsabs.harvard.edu/abs/2015PhRvD..91d3002L} {91, 043002}

\bibitem[\protect\citeauthoryear{{Landry} \& {Essick}}{{Landry} \&
  {Essick}}{2019}]{landry19}
{Landry} P.,  {Essick} R.,  2019, \mn@doi [\prd] {10.1103/PhysRevD.99.084049},
  \href {https://ui.adsabs.harvard.edu/abs/2019PhRvD..99h4049L} {99, 084049}

\bibitem[\protect\citeauthoryear{{Legred}, {Chatziioannou}, {Essick}, {Han}  \&
  {Landry}}{{Legred} et~al.}{2021}]{legred21}
{Legred} I.,  {Chatziioannou} K.,  {Essick} R.,  {Han} S.,   {Landry} P.,
  2021, \mn@doi [\prd] {10.1103/PhysRevD.104.063003}, \href
  {https://ui.adsabs.harvard.edu/abs/2021PhRvD.104f3003L} {104, 063003}

\bibitem[\protect\citeauthoryear{{Maggiore} et~al.}{{Maggiore}
  et~al.}{2020}]{maggiore20}
{Maggiore} M.,  et~al., 2020, \mn@doi [\jcap] {10.1088/1475-7516/2020/03/050},
  \href {https://ui.adsabs.harvard.edu/abs/2020JCAP...03..050M} {2020, 050}

\bibitem[\protect\citeauthoryear{{Mandel}, {Farr}  \& {Gair}}{{Mandel}
  et~al.}{2019}]{mandel19}
{Mandel} I.,  {Farr} W.~M.,   {Gair} J.~R.,  2019, \mn@doi [\mnras]
  {10.1093/mnras/stz896}, \href
  {https://ui.adsabs.harvard.edu/abs/2019MNRAS.486.1086M} {486, 1086}

\bibitem[\protect\citeauthoryear{{Margalit} \& {Metzger}}{{Margalit} \&
  {Metzger}}{2017}]{margalit17}
{Margalit} B.,  {Metzger} B.~D.,  2017, \mn@doi [\apjl]
  {10.3847/2041-8213/aa991c}, \href
  {https://ui.adsabs.harvard.edu/abs/2017ApJ...850L..19M} {850, L19}

\bibitem[\protect\citeauthoryear{{Masuda}, {Hatsuda}  \& {Takatsuka}}{{Masuda}
  et~al.}{2016}]{masuda16}
{Masuda} K.,  {Hatsuda} T.,   {Takatsuka} T.,  2016, \mn@doi [European Physical
  Journal A] {10.1140/epja/i2016-16065-6}, \href
  {https://ui.adsabs.harvard.edu/abs/2016EPJA...52...65M} {52, 65}

\bibitem[\protect\citeauthoryear{{Miller} et~al.,}{{Miller}
  et~al.}{2019}]{miller19}
{Miller} M.~C.,  et~al., 2019, \mn@doi [\apjl] {10.3847/2041-8213/ab50c5},
  \href {https://ui.adsabs.harvard.edu/abs/2019ApJ...887L..24M} {887, L24}

\bibitem[\protect\citeauthoryear{{Miller} et~al.,}{{Miller}
  et~al.}{2021}]{miller21}
{Miller} M.~C.,  et~al., 2021, \mn@doi [\apjl] {10.3847/2041-8213/ac089b},
  \href {https://ui.adsabs.harvard.edu/abs/2021ApJ...918L..28M} {918, L28}

\bibitem[\protect\citeauthoryear{{Most}, {Weih}, {Rezzolla}  \&
  {Schaffner-Bielich}}{{Most} et~al.}{2018}]{most18}
{Most} E.~R.,  {Weih} L.~R.,  {Rezzolla} L.,   {Schaffner-Bielich} J.,  2018,
  \mn@doi [\prl] {10.1103/PhysRevLett.120.261103}, \href
  {https://ui.adsabs.harvard.edu/abs/2018PhRvL.120z1103M} {120, 261103}

\bibitem[\protect\citeauthoryear{{Most}, {Papenfort}, {Dexheimer}, {Hanauske},
  {Schramm}, {St{\"o}cker}  \& {Rezzolla}}{{Most} et~al.}{2019}]{most19}
{Most} E.~R.,  {Papenfort} L.~J.,  {Dexheimer} V.,  {Hanauske} M.,  {Schramm}
  S.,  {St{\"o}cker} H.,   {Rezzolla} L.,  2019, \mn@doi [\prl]
  {10.1103/PhysRevLett.122.061101}, \href
  {https://ui.adsabs.harvard.edu/abs/2019PhRvL.122f1101M} {122, 061101}

\bibitem[\protect\citeauthoryear{{Pang}, {Dietrich}, {Tews}  \& {Van Den
  Broeck}}{{Pang} et~al.}{2020}]{pang20}
{Pang} P. T.~H.,  {Dietrich} T.,  {Tews} I.,   {Van Den Broeck} C.,  2020,
  \mn@doi [Physical Review Research] {10.1103/PhysRevResearch.2.033514}, \href
  {https://ui.adsabs.harvard.edu/abs/2020PhRvR...2c3514P} {2, 033514}

\bibitem[\protect\citeauthoryear{{Punturo} et~al.}{{Punturo}
  et~al.}{2010}]{punturo10}
{Punturo} M.,  et~al., 2010, \mn@doi [Classical and Quantum Gravity]
  {10.1088/0264-9381/27/19/194002}, \href
  {https://ui.adsabs.harvard.edu/abs/2010CQGra..27s4002P} {27, 194002}

\bibitem[\protect\citeauthoryear{{Raaijmakers} et~al.,}{{Raaijmakers}
  et~al.}{2021}]{raaijmakers21}
{Raaijmakers} G.,  et~al., 2021, \mn@doi [\apjl] {10.3847/2041-8213/ac089a},
  \href {https://ui.adsabs.harvard.edu/abs/2021ApJ...918L..29R} {918, L29}

\bibitem[\protect\citeauthoryear{{Radice}, {Perego}, {Zappa}  \&
  {Bernuzzi}}{{Radice} et~al.}{2018}]{radice18}
{Radice} D.,  {Perego} A.,  {Zappa} F.,   {Bernuzzi} S.,  2018, \mn@doi [\apjl]
  {10.3847/2041-8213/aaa402}, \href
  {https://ui.adsabs.harvard.edu/abs/2018ApJ...852L..29R} {852, L29}

\bibitem[\protect\citeauthoryear{{Riley} et~al.,}{{Riley}
  et~al.}{2019}]{riley19}
{Riley} T.~E.,  et~al., 2019, \mn@doi [\apjl] {10.3847/2041-8213/ab481c}, \href
  {https://ui.adsabs.harvard.edu/abs/2019ApJ...887L..21R} {887, L21}

\bibitem[\protect\citeauthoryear{{Riley} et~al.,}{{Riley}
  et~al.}{2021}]{riley21}
{Riley} T.~E.,  et~al., 2021, \mn@doi [\apjl] {10.3847/2041-8213/ac0a81}, \href
  {https://ui.adsabs.harvard.edu/abs/2021ApJ...918L..27R} {918, L27}

\bibitem[\protect\citeauthoryear{{Romero-Shaw} et~al.,}{{Romero-Shaw}
  et~al.}{2020}]{romeroshaw20}
{Romero-Shaw} I.~M.,  et~al., 2020, \mn@doi [\mnras] {10.1093/mnras/staa2850},
  \href {https://ui.adsabs.harvard.edu/abs/2020MNRAS.499.3295R} {499, 3295}

\bibitem[\protect\citeauthoryear{{Sarin} \& {Lasky}}{{Sarin} \&
  {Lasky}}{2021}]{sarin21}
{Sarin} N.,  {Lasky} P.~D.,  2021, \mn@doi [General Relativity and Gravitation]
  {10.1007/s10714-021-02831-1}, \href
  {https://ui.adsabs.harvard.edu/abs/2021GReGr..53...59S} {53, 59}

\bibitem[\protect\citeauthoryear{{Sarin}, {Lasky}  \& {Ashton}}{{Sarin}
  et~al.}{2020}]{sarin20}
{Sarin} N.,  {Lasky} P.~D.,   {Ashton} G.,  2020, \mn@doi [\prd]
  {10.1103/PhysRevD.101.063021}, \href
  {https://ui.adsabs.harvard.edu/abs/2020PhRvD.101f3021S} {101, 063021}

\bibitem[\protect\citeauthoryear{{Shirke}, {Ghosh}  \& {Chatterjee}}{{Shirke}
  et~al.}{2023}]{shirke23}
{Shirke} S.,  {Ghosh} S.,   {Chatterjee} D.,  2023, \mn@doi [\apj]
  {10.3847/1538-4357/acac31}, \href
  {https://ui.adsabs.harvard.edu/abs/2023ApJ...944....7S} {944, 7}

\bibitem[\protect\citeauthoryear{{Smith}, {Field}, {Blackburn}, {Haster},
  {P{\"u}rrer}, {Raymond}  \& {Schmidt}}{{Smith} et~al.}{2016}]{smith16}
{Smith} R.,  {Field} S.~E.,  {Blackburn} K.,  {Haster} C.-J.,  {P{\"u}rrer} M.,
   {Raymond} V.,   {Schmidt} P.,  2016, \mn@doi [\prd]
  {10.1103/PhysRevD.94.044031}, \href
  {https://ui.adsabs.harvard.edu/abs/2016PhRvD..94d4031S} {94, 044031}

\bibitem[\protect\citeauthoryear{{Sun}, {Kuns}, {Slagmolen}  et~al.}{{Sun}
  et~al.}{2026}]{sun26}
{Sun} L.,  {Kuns} K.,  {Slagmolen} B.~J.~J.,   et~al., 2026, arXiv e-prints,
  \href {https://ui.adsabs.harvard.edu/abs/2026arXiv260811673S} {p.
  arXiv:2608.11673}

\bibitem[\protect\citeauthoryear{{Takami}, {Rezzolla}  \& {Baiotti}}{{Takami}
  et~al.}{2014}]{takami14}
{Takami} K.,  {Rezzolla} L.,   {Baiotti} L.,  2014, \mn@doi [\prl]
  {10.1103/PhysRevLett.113.091104}, \href
  {https://ui.adsabs.harvard.edu/abs/2014PhRvL.113i1104T} {113, 091104}

\bibitem[\protect\citeauthoryear{{Thrane} \& {Talbot}}{{Thrane} \&
  {Talbot}}{2019}]{thrane19}
{Thrane} E.,  {Talbot} C.,  2019, \mn@doi [\pasa] {10.1017/pasa.2019.2}, \href
  {https://ui.adsabs.harvard.edu/abs/2019PASA...36...10T} {36, e010}

\bibitem[\protect\citeauthoryear{{Typel}, {Oertel}  \& {Kl{\"a}hn}}{{Typel}
  et~al.}{2015}]{typel15}
{Typel} S.,  {Oertel} M.,   {Kl{\"a}hn} T.,  2015, \mn@doi [Physics of
  Particles and Nuclei] {10.1134/S1063779615040061}, \href
  {https://ui.adsabs.harvard.edu/abs/2015PPN....46..633T} {46, 633}

\bibitem[\protect\citeauthoryear{{Typel} et~al.,}{{Typel}
  et~al.}{2022}]{typel22}
{Typel} S.,  et~al., 2022, \mn@doi [European Physical Journal A]
  {10.1140/epja/s10050-022-00847-y}, \href
  {https://ui.adsabs.harvard.edu/abs/2022EPJA...58..221C} {58, 221}

\bibitem[\protect\citeauthoryear{{Weih}, {Hanauske}  \& {Rezzolla}}{{Weih}
  et~al.}{2020}]{weih20}
{Weih} L.~R.,  {Hanauske} M.,   {Rezzolla} L.,  2020, \mn@doi [\prl]
  {10.1103/PhysRevLett.124.171103}, \href
  {https://ui.adsabs.harvard.edu/abs/2020PhRvL.124q1103W} {124, 171103}

\makeatother
\end{thebibliography}

\end{document}